\documentclass[reprint,onecolumn,superscriptaddress,a4paper,nofootinbib,floatfix]{revtex4-1}
\usepackage[utf8]{inputenc}
\usepackage{graphicx}
\usepackage{footnote}
\usepackage{bm}
\usepackage{microtype}
\usepackage{amsmath}
\usepackage{amsfonts} 
\usepackage{subfig}
\usepackage{textcomp}
\usepackage{dcolumn}
\usepackage{amssymb}
\usepackage{geometry}
\usepackage{subfig}
\usepackage{color}
\usepackage{microtype}
\usepackage{graphicx}
\usepackage{footnote}
\usepackage{microtype}
\usepackage{amsmath}
\usepackage{amsfonts} 
\usepackage{amssymb}
\usepackage{geometry}
\usepackage{textcomp}
\usepackage{mwe}
\usepackage[section]{placeins}
\usepackage[font={small}]{caption}

\begin{document}
\title{Dimensional crossover and incipient quantum size effects in superconducting niobium nanofilms}

\author{N. Pinto}
\affiliation{School of Science and Technology, Physics division, University of Camerino, Camerino, Italy}
\author{S. Javad Rezvani}
\affiliation{School of Science and Technology, Physics division, University of Camerino, Camerino, Italy}
\affiliation{Istituto Nazionale di Ricerca Metrologica (INRiM), Strada delle Cacce, 91, Torino, 10135, Italy}
\author{Andrea Perali}
\affiliation{School of Pharmacy, Physics Unit, University of Camerino, Camerino, 62032, Italy}
\author{Luca Flammia}
\affiliation{School of Science and Technology, Physics division, University of Camerino, Camerino, Italy}
\author{Milorad V. Milo\v{s}evi\'{c}}
\affiliation{Department of Physics, University of Antwerp, Groenenborgerlaan 171, B-2020 Antwerp, Belgium}
\author{Matteo Fretto}
\affiliation{Istituto Nazionale di Ricerca Metrologica (INRiM), Strada delle Cacce, 91, Torino, 10135, Italy}
\author{Cristina Cassiago}
\affiliation{Istituto Nazionale di Ricerca Metrologica (INRiM), Strada delle Cacce, 91, Torino, 10135, Italy}
\author{Natascia De Leo}
\affiliation{Istituto Nazionale di Ricerca Metrologica (INRiM), Strada delle Cacce, 91, Torino, 10135, Italy}

\keywords{Niobium, nanofilms, superconductivity, transport properties, quantum confinement}

\begin{abstract}
Superconducting and normal state properties of sputtered Niobium nanofilms have been systematically investigated, as a function of film thickness in a \textit{d}=9-90 nm range, on different substrates. The width of the superconducting-to-normal transition for all films remained in few tens of mK, thus remarkably narrow, confirming their high quality. We found that the superconducting critical current density exhibits a pronounced maximum, three times larger than its bulk value, for film thickness around 25 nm, marking the 3D-to-2D crossover. The extracted magnetic penetration depth shows a sizeable enhancement for the thinnest films, aside the usual demagnetization effects. Additional amplification effects of the superconducting properties have been obtained in the case of sapphire substrates or squeezing the lateral size of the nanofilms. For thickness close to 20 nm we also measured a doubled perpendicular critical magnetic field compared to its saturation value for \textit{d}$>$33 nm, indicating shortening of the correlation length and the formation of small Cooper pairs in the condensate. Our data analysis evidences an exciting interplay between quantum-size and proximity effects together with strong-coupling effects and importance of disorder in the thinnest films, locating the ones with optimally enhanced critical properties close to the BCS-BEC crossover regime.
\end{abstract}

\date{\today}
%\begin{document}
\flushbottom
\maketitle

\section*{Introduction}
Superconducting niobium thin films are currently investigated for both fundamental and technological applications.
Thin films of Nb with a thickness of a few tens or even few nanometers, grown on suitable substrates are needed for technological quantum devices, such as Josephson junctions, nano Superconducting QUantum Interference Devices (SQUIDs), \cite{Russo:2014} mixers \cite{Karasik} and single photon detectors. \cite {Semenov:2003}

The fabrication and characterization of high quality Nb nanofilms with optimized superconducting properties is of key importance. 
Indeed, these layers can be the ideal platform to design nanostructures with reduced dimensionality, such as single or multiple superconducting nano-stripes, to control and enhance superconductivity; \cite{Perali96,Bianconi97,Bianconi98,Perali2012} to generate multigap and resonant phenomena; \cite{Blatt1,Blatt2,Innocenti2010,PeraliSUST,2DEGPerali,MMAP2015} to fabricate heterostructures of superconductors with other materials, \cite{Buzdin2005} etc. A plethora of those and other novel nanostructures can be fabricated starting from Nb nanofilms, using readily available techniques of, e.g., electron beam lithography and focused ion beam nano-sculpting. \cite{Fretto2013}

In this work we report an extended experimental investigation of the superconducting to normal state transition in Nb nanofilms, with a thickness in the range 9 nm to  90 nm. All the Nb nanofilms on different substrates have been fabricated by sputtering technique, as described in the Methods section.
We have measured and derived the main parameters characterising the normal and the superconducting state, such as the mean free path of charge carriers, the critical current, the magnetic penetration depth and the correlation length (related to the average size of the Cooper pairs). Their values have been analysed and discussed in terms of the recent theories evidencing, in some cases, the limits of the available models. 
We have also explored the important role of the substrate, considering Nb nanofilms grown on two different substrates: silicon dioxide (SiO$_2$) and sapphire (Al$_ 2$O$_3$).
They form quite different interfaces with Nb, in terms of oxidation of the grown nanofilms. In addition, having as a final goal the fabrication of superconducting nanostripes, we started
in this work to investigate the effect of the lateral dimension of the nanofilms, considering samples of different widths.

Our work delivers also evidences for a 3D to 2D dimensional crossover in the superconducting properties of the Nb nanofilms as the thickness is reduced below $\approx 20\div 30$ nm. The dimensional crossover is deduced from the observation of abrupt changes in the thickness dependence of the critical temperature, the critical current and the superconducting length scales.  We also discuss the behavior of the critical temperature as a function of the film thickness in terms of an interplay between quantum size effects and proximity effects with the oxide formed at the interface with the substrate, in the superconducting phase of the nanofilms. Additionally, we have provided evidence for Cooper-pair shrinking when approaching the ultra thin regime of film thicknesses, that is discussed within the framework of the Bardeen-Cooper-Schrieffer (BCS) - Bose-Einstein Condensation (BEC) crossover theory, according to the prediction for molecular pairing in superconducting nanofilms.\cite{Chen2012} We have demonstrated that the thickness dependence of the mean free path is responsible for the Cooper pair shrinking due to non-local effects and for the large variations of the magnetic penetration depth as a function of thickness. This result further indicates the non trivial interplay between disorder and BCS-BEC crossover. It is already known that, in general on the BCS weak coupling side of the BCS-BEC crossover, disorder increases the superconducting critical temperature and amplifies the superconducting fluctuations responsible for the pseudogap opening in the single-particle excitation spectrum.\cite{Palestini2013} 

Finally, we discuss how the comprehensive experimental analysis of the superconducting properties of Nb nanofilms, together with the theoretical interpretation presented in this work, gives clear indication on the optimal performance of the Nb films for thickness between 20 nm and 30 nm.
From a technological point of view, these outcomes fill the gap of information on superconducting properties of the films, between the atomistic and the mesoscopic range of thickness, and will be useful to guide the fabrication of optimized nanodevices and other superconducting circuitry.
 
\section*{Results}
\subsection*{Resistivity and critical temperature}
\begin{figure}[ht]
\centering
\includegraphics[width=\linewidth]{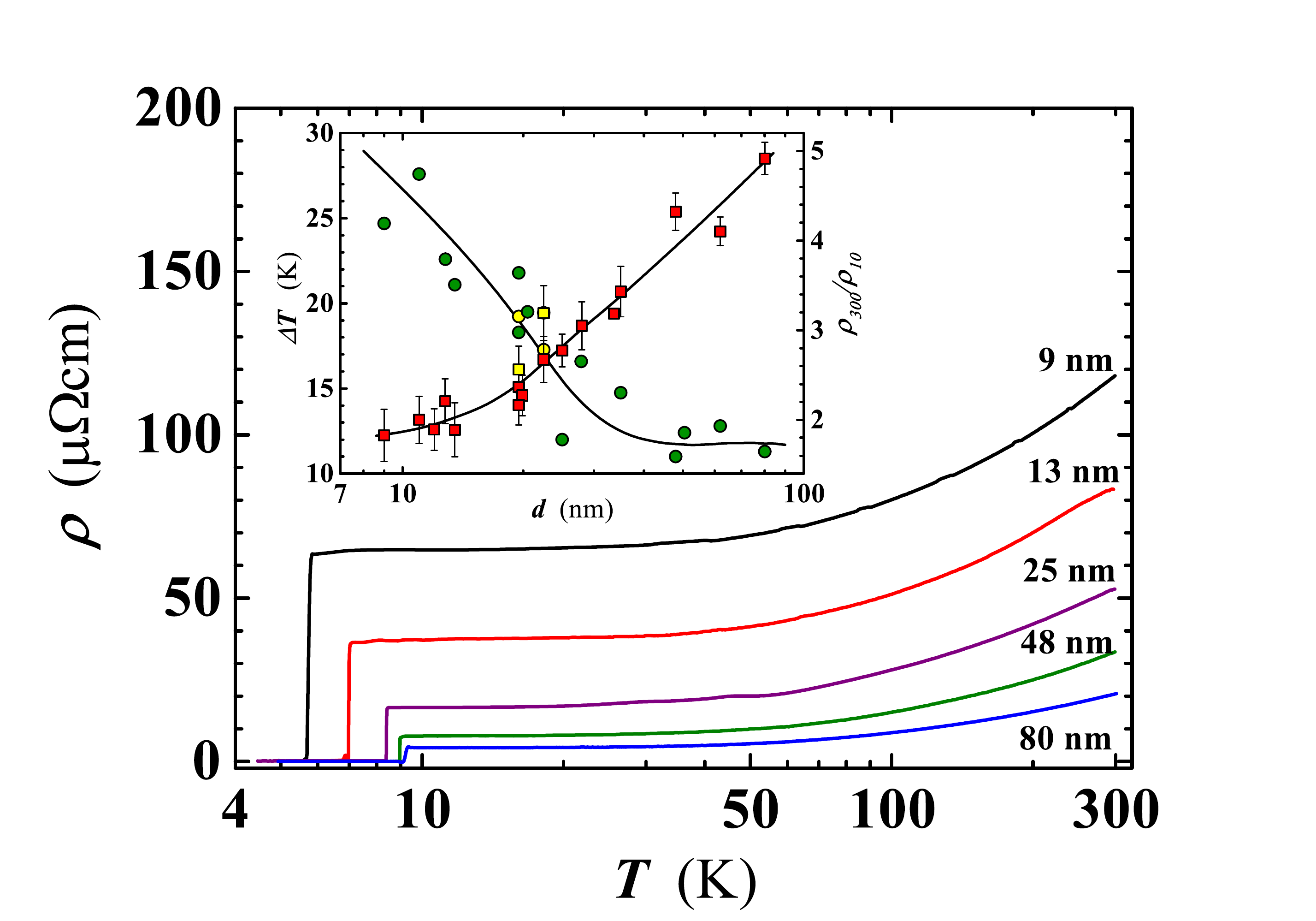}
\caption{Temperature dependence of the resistivity for selected Nb films, with a thickness ranging from 9 nm to 80 nm. Inset: width of the plateau region ($\Delta T$), above $T_C$, as a function of $d$ (circles). As $\Delta T$ we have considered the $T$ interval corresponding to a film resistivity change of $\pm 2.5$\% of $\rho(T)$ @ 10 K. Squares: resistivity ratio at $T = 300$ K and at $T = 10$ K as a function of $d$. Yellow filled symbols refer to layers deposited on sapphire substrates. Lines serve as a guide for the eyes.}  
\label{FigRT}
\end{figure}

The temperature dependence of the resistivity, $\rho(T)$, of Nb films has been investigated as a function of the thickness, $d$, in the range $9 \lesssim d \lesssim 80$ nm (see Table \ref{Table}).\footnote{For details on fabrication of the films, we refer the reader to the Methods section.} 
The $\rho(T)$ curves shift upward, while the superconducting transition temperature, $T_C$, decreases as $d$ is progressively reduced (Figure \ref{FigRT}). We have found that for films with $9 < d < 25$ nm, the shift of the $\rho(T)$ curves is larger than for $25 \leq d < 80$ nm (Figure \ref{FigRT}). 
The 80 nm thick film ($\#6$), being least resistive of those measured, exhibits a room temperature (R.T.) value of $\rho(T) \approx 20$ $\mu\Omega$cm, approaching the Nb bulk value (15 $\mu\Omega$cm).\cite{Bulk_res1,Bulk_res2} Lowering of the resistivity with thicker $d$ suggests a gradual reduction of the film defectivity, in accordance with the behavior reported for Nb films by other groups.\cite{Zhao,Delacour,Gubin2005}
For $T_C < T \lesssim 45$ K, the $\rho(T)$ shows a plateau, due to the residual defects in the nanofilm (the nature and the influence of which will be discussed later). 
Starting from the thinnest films, this plateau rapidly shrinks with increasing $d$ up to 25 nm and then attains a nearly constant value for $25 < d < 80$ nm (see inset of Figure \ref{FigRT}). 

\begin{figure}[h]
\centering
\includegraphics[width=\linewidth]{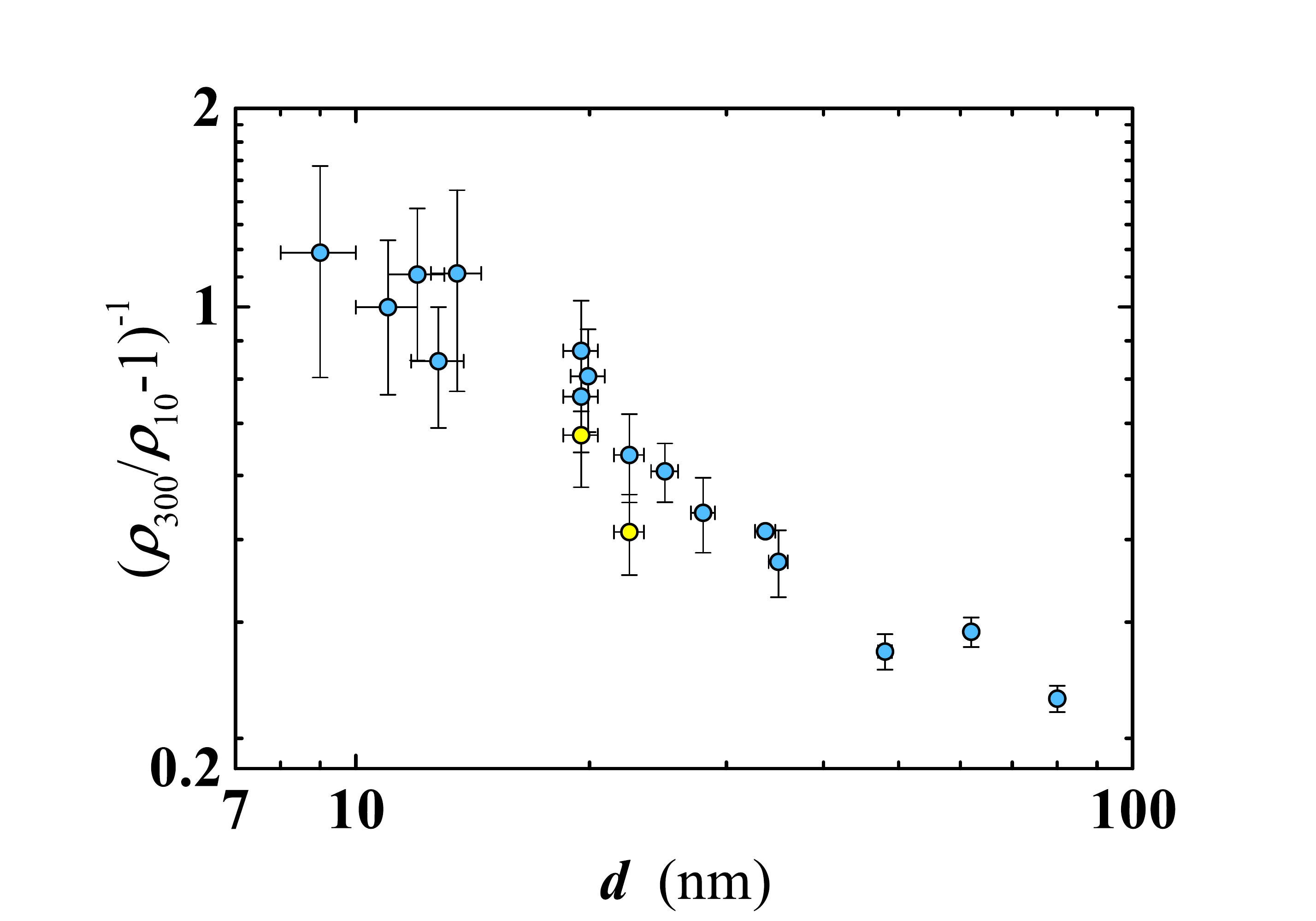}
\caption{Thickness dependence of the quantity $C = (\rho_{300}/\rho_{10}-1)^{-1}$. $\rho_{300}/\rho_{10}$ is the residual resistivity ratio (i.e. $RRR$). All films have been deposited on silicon substrates except for those indicated by yellow filled circles, deposited on sapphire. } 
\label{RRR}
\end{figure}

The gradual improvement of the film quality for larger $d$ is also deduced by the thickness dependence both of the residual resistivity ratio, $RRR$, here defined as the $\rho(T)$ ratio at $T = 300$ K and $T= 10$ K, i.e. $RRR = \rho_{300}/\rho_{10}$\cite{Lotnyk2016} and the quantity $C = (RRR-1)^{-1}$, proportional to defects density in the film.\cite{Delacour} 
$RRR$ grows monotonically with $d$, from $\sim 2$, for $d = 9\div13$ nm, to $\sim 5$ at $d = 80$ nm (Figure \ref{FigRT}: inset). These values are comparable with the data reported by Lotnyk \textit{et al.}\cite{Lotnyk2016} and, for $d < 40$ nm, also with data of Mayadas \textit{et al.}\cite{Mayadas1972} 
Agreement with findings of Delacour \textit{et al.}\cite{Delacour} occurs at the lowest thickness ($d \approx 10$ nm). 
Finally, the rapid lowering of $C$ with increasing film thickness indicates reduction of the defect density (Figure \ref{RRR}), in agreement with findings of Ref. \cite{Delacour}. It is worth noting that $C$ becomes lower, for the same thickness, in those films deposited on the sapphire substrate (see Figure \ref{RRR}). Possible effects of the substrate on other measured film properties will be addressed later.

\begin{figure}[h]
\centering
\includegraphics[width=\linewidth]{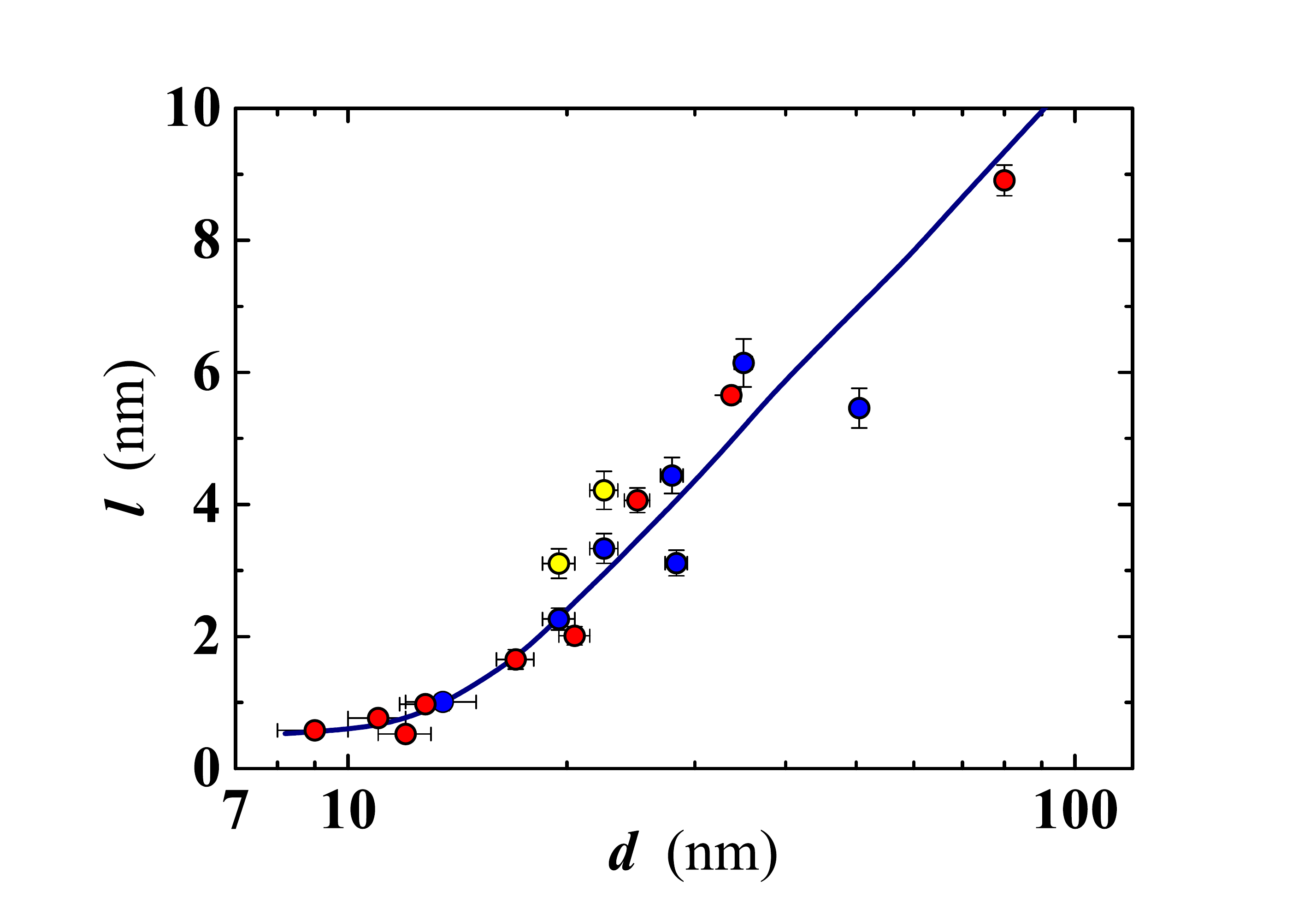}
\caption{Thickness dependence of the mean free path at 10 K. Data have been derived from the resistivity values measured at the same temperature (i.e. $\rho_{10}$) for either 10 $\mu$m (blue dots) or 50 $\mu$m (red dots) wide films. Yellow filled symbols correspond to films (10 $\mu$m wide) deposited on a sapphire substrate; all the other films have been deposited on silicon. Continuous line is a guide for eyes.} 
\label{Lmean}
\end{figure}

Residual resistivity (i.e. $\rho_{10}$) can be used to estimate the charge carriers mean-free path, $l$, at $T = 10$ K, assuming the constancy of the product $\rho l = 3.75 \times 10^{-6}$ $\mu\Omega$cm$^2$, $\rho$ being the bulk Nb resistivity.\cite{Ashcroft} Values for $l$ range from $\approx 1$ nm, at the lower thicknesses, to $\simeq 9$ nm at $d = 80$ nm (Figure \ref{Lmean}). Films deposited on sapphire evidence higher $l$ values, in agreement with an expected lower defect density in the film matrix (Figure \ref{Lmean}).
Further on in our analysis, the thickness dependence of $l$ will be used to extract the thickness dependence of superconducting coherence length and magnetic penetration depth.

\begin{figure}[ht]
\centering
\includegraphics[width=\linewidth]{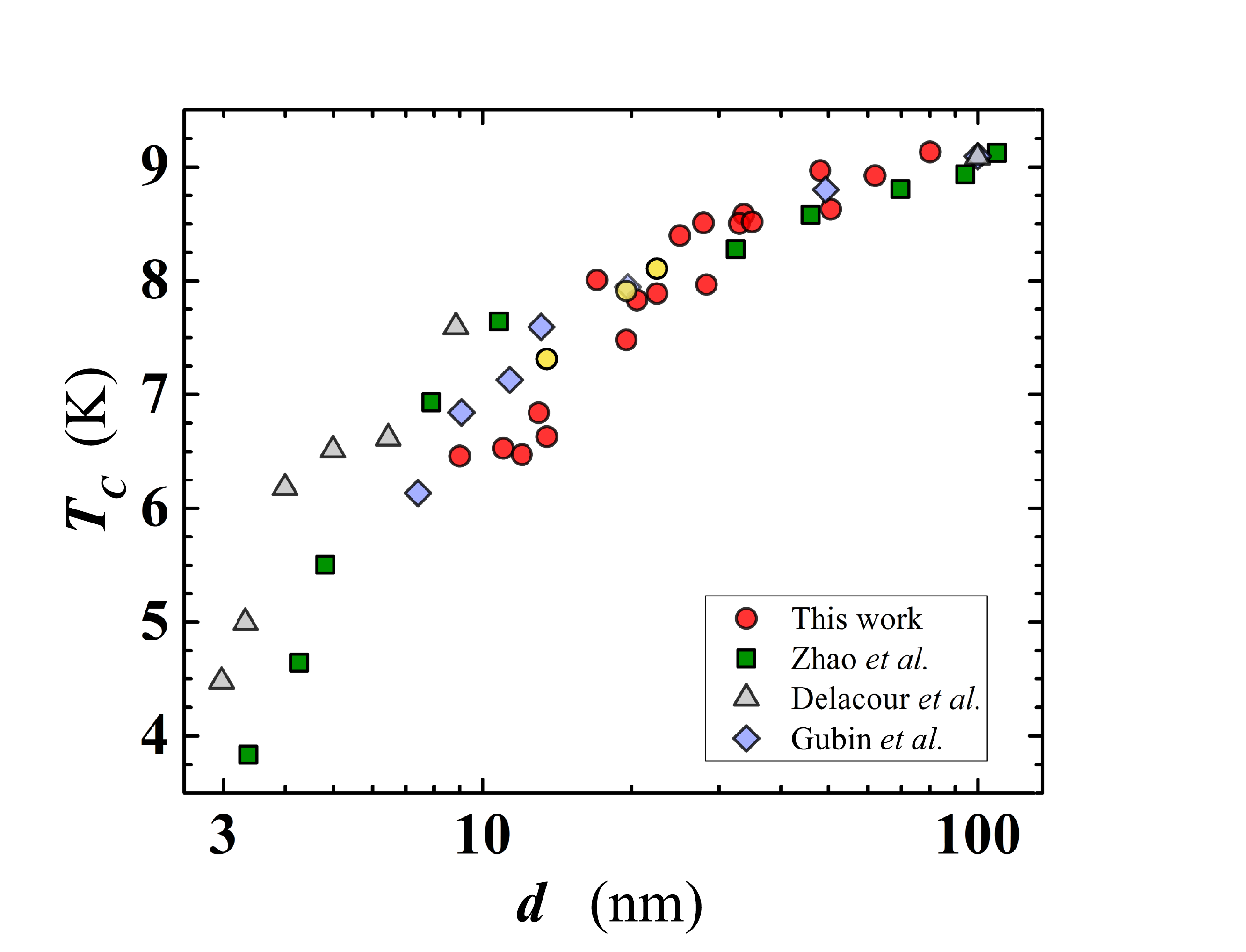}
\caption{Thickness dependence of the superconducting transition temperature. Circles: data of the present work, for layers deposited on SiO$_2$ (red filled) and on sapphire (yellow filled). For comparison, we have added data from: Zhao \textit{et al.} (squares) \cite{Zhao}; Gubin \textit{et al.} (diamonds) \cite{Gubin2005}; Delacour \textit{et al.} (triangles).\cite{Delacour} }  
\label{FigTc}
\end{figure}

\begin{figure}[ht]
\centering
\includegraphics[width=\linewidth]{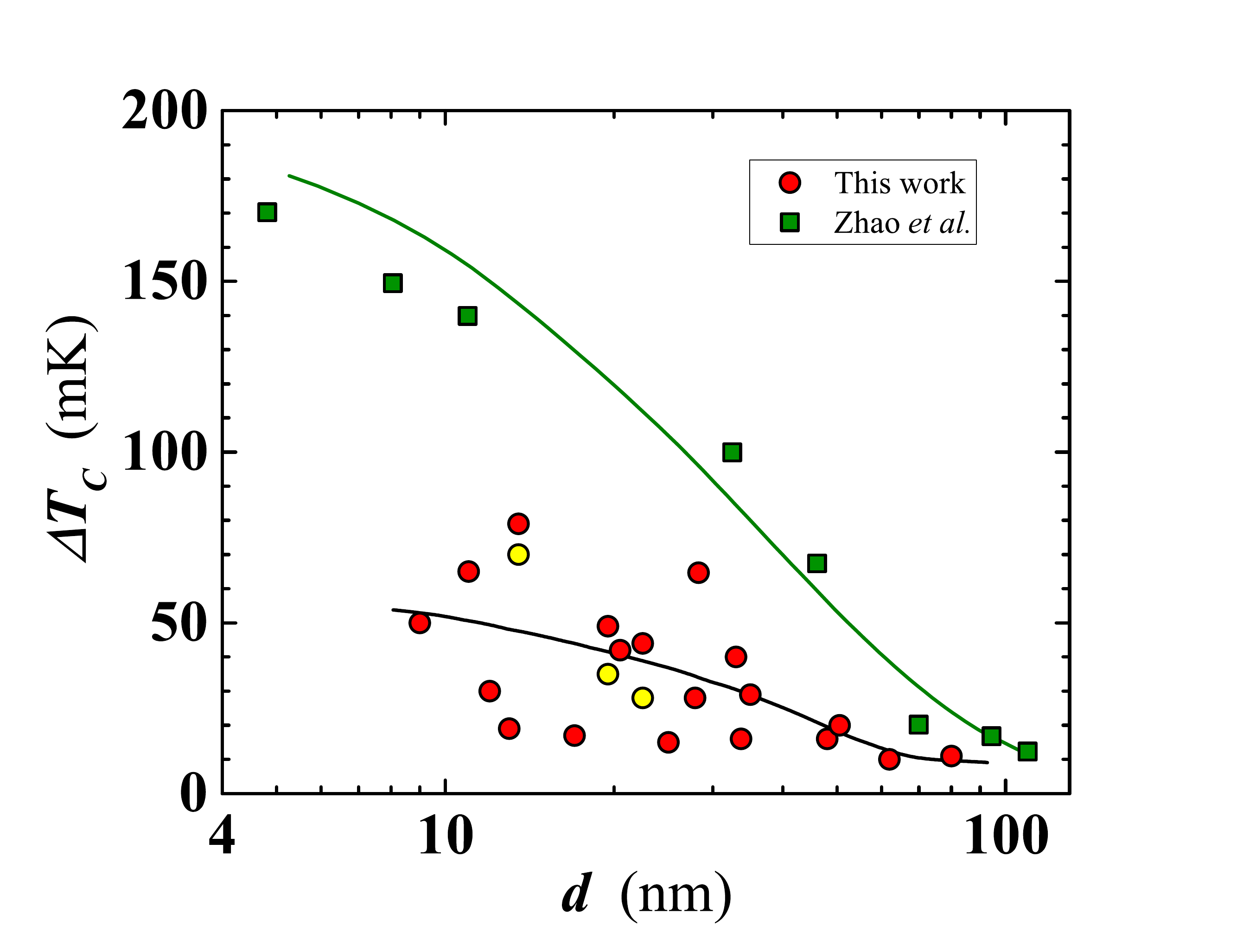}
\caption{Thickness dependence of the superconducting transition width. Circles: data of the present work, measured in films deposited on SiO$_2$ and on sapphire (yellow filled). Squares: data taken from Zhao \textit{et al.} \cite{Zhao} Solid lines serve as a guide for the eyes.}  
\label{DeltaTc_vs_d}
\end{figure}

Next, we have studied in detail the dependence of the superconducting transition temperature on film thickness. We have determined $T_C$ as the midpoint between temperatures corresponding to the 10\% and the 90\% of the normal resistance value, located on the residual resistance plateau of the $R(T)$ curve, for $T > T_C$. The difference of those two $T$ values was taken as the width of the superconducting transition, $\Delta T_C$. As expected, $T_C$ decreases with decreasing thickness, from $T_C \simeq 9.1$ K for $d = 80$ nm to $T_C \simeq 6.5$ K for $\approx 10$ nm. For $d$ values lower than $\simeq 13$ nm, $T_C$ decreases more abruptly, with similar $T_C$ values found for our five thinnest films (Figure \ref{FigTc}). Compared to $T_C$ data reported in the literature by other groups, the $T_C$ of our films is somewhat higher for $d$ larger than 17 nm, becoming lower for smaller thicknesses (Figure \ref{FigTc}). This suggests certain level of disorder in our thinnest films. To check that, we examined the width of the superconducting transition, $\Delta T_C$, measured as $15\div30$ mK, for $25 \leq d < 80$ nm, increasing to $\simeq 70 \div 80$ mK for the smallest thicknesses ($d < 25$ nm; Figure \ref{DeltaTc_vs_d}). Zhao \textit{et al.} \cite{Zhao} reported a rapidly rising $\Delta T_C$ for $d < 30$ nm, with multiply larger values than those measured in our films with comparable thickness (Figure \ref{DeltaTc_vs_d}). These results suggest a superior quality of our superconducting Nb nanofilms at all thicknesses. 

What is then the reason for observed lower $T_C$ values for $d < 17$ nm than was the case in Ref. \cite{Zhao} ? This is likely an unwanted contribution from an oxidized Nb (i.e. NbO$_x$) layer \cite{Gershenzon}, formed at the interface between the Nb film and the SiO$_2$ substrate, becoming progressively stronger with the reduction of $d$. This oxide layer decreases the quality of the film-substrate interface, and reduces the effective Nb film thickness. Such a contribution is largely reduced on sapphire substrates, which could explain the smoother decrease of $T_C(d)$ detected by Zhao \textit{et al.} \cite{Zhao}. To verify this, we carried out a specific check, depositing 13.5 nm, 19.5 nm and 22.5 nm thick Nb films on a sapphire substrate (i.e. $\#H14S$, $\#H1S$ and $\#H11S$, respectively). Indeed, in those films $T_C$ was significantly higher ($\sim 0.5$ K for $d = 13.5$ nm, see yellow filled circles in Figure \ref{FigTc}) than in those deposited on a SiO$_2$ substrate, obtained during the same run ($\#H15$ \& $\#H1$; $\#H9$: see Table I). The reduction of $C$ and the increase of $l$, in films deposited on sapphire, is in agreement with the hypothesis of a better quality of the layer at the film-substrate interface.

\begin{figure}[ht]
\centering
\includegraphics[width=\linewidth]{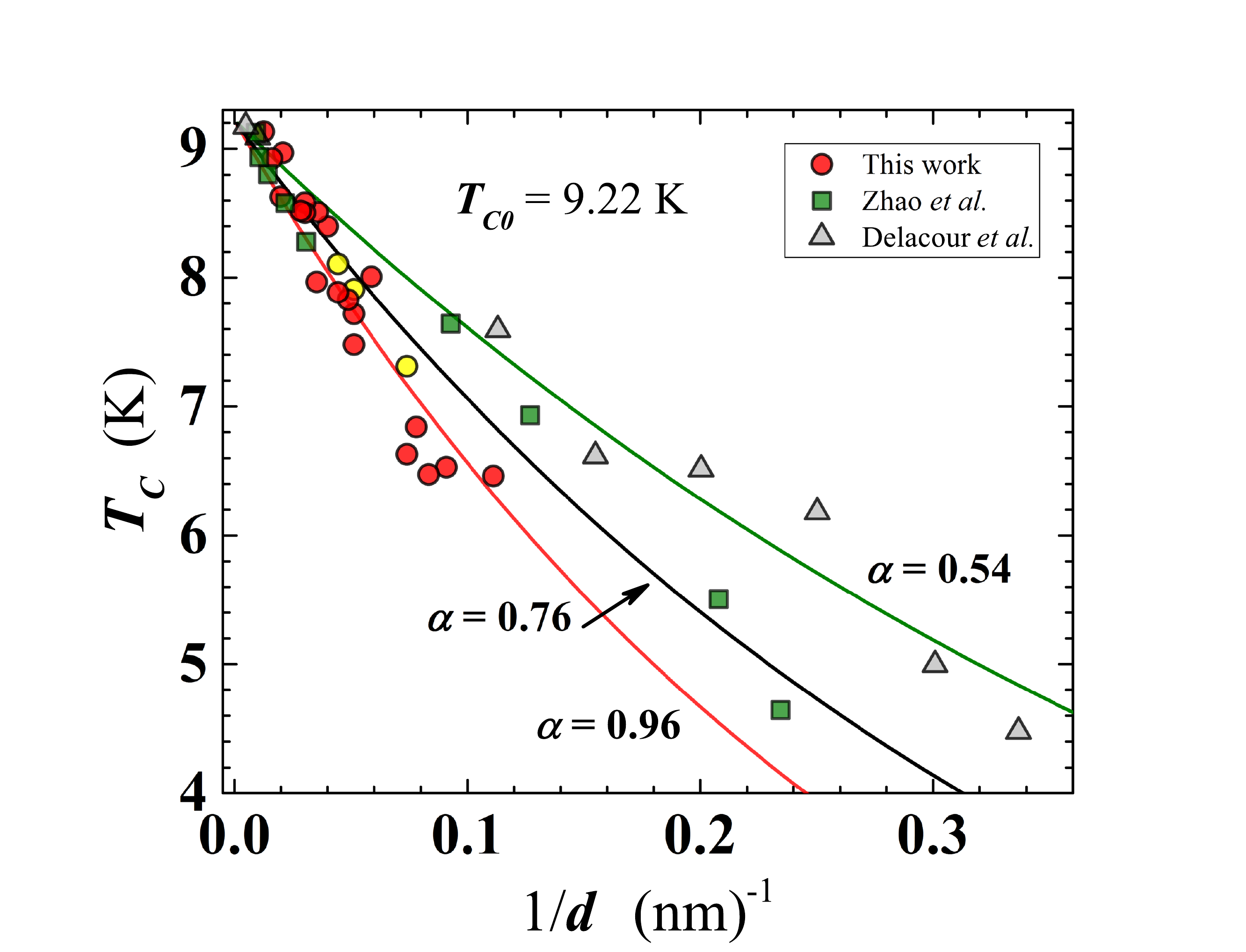}
\caption{Superconducting transition temperature as a function of the reciprocal film thickness. Present work (SiO$_2$ substrate: red filled circles; sapphire substrate: yellow filled circles); Zhao \textit{et al.}\cite{Zhao} (triangles); Delacour \textit{et al.}\cite{Delacour} (squares). Continuous lines are least-squares fit, using Eq. \eqref{Tcantiprox} (see text), of the different data sets. The effective thickness of the normal layer (in nm) is represented by the corresponding $\alpha$ value for each curve.}  
\label{fit_Tc_vs_d-1}
\end{figure}

Another likely effect of the formed NbO$_x$ layer is that, being a conductive system of electrons in their normal state, it is able to sink Cooper pairs from the superconducting Nb nanofilm, suppressing the condensate fraction and reducing the $T_C$ via proximity effect. This phenomenon should become more effective as the thickness of the Nb nanofilm is reduced so that NbO$_x$ layer becomes a sizeable fraction of the Nb film thickness. Indeed, the $T_C(d)$ curve of Figure \ref{FigTc} shows a smooth decrease of $T_C$ for intermediate-to-large film thickness (i.e. for $20 \lesssim d \lesssim 80$ nm), which accelerates for smaller $d$ values (i.e. at $d \lesssim 20$ nm). The $T_C$ suppression law, due to the proximity effect, has been derived by McMillan\cite{McMillan} and is given by:
\begin{equation}
\centering
T_C = T_{C0}\left( \frac{3.56 T_D}{T_{C0}\pi} \right)^{-\alpha/d},
\label{Tcantiprox}
\end{equation}
where $\alpha = d_N N_N(0)/N_S(0)$ is an effective thickness of the conductive layer at the interface; $T_{C0} = 9.22$ K is the bulk $T_C$ of Nb and $T_D$ = 277 K is the Debye temperature. The quantities $N_N(0)$ and $N_S(0)$ are the density of states in the normal ($N$) and superconducting ($S$) layers, respectively. For simplicity, considering the ratio $N_N (0)/N_S (0)$ equal to unity, one obtains $\alpha \cong d_N$. Delacour \textit{et al.} \cite{Delacour} applied Eq. \eqref{Tcantiprox} to fit their data set for $T_C(d)$ of Nb films on sapphire, finding a very small effective thickness of the normal layer of $d_N = 0.54 \pm 0.02$ nm. Following the same procedure and interpretation of Delacour \textit{et al.}, we obtain a satisfactory fit of our data for $T_C(d)$ using Eq. \eqref{Tcantiprox} and $d_N = 0.96\pm 0.04$ nm (Figure \ref{fit_Tc_vs_d-1}). In addition, we have found an intermediate value of $d_N = 0.76\pm 0.03$ nm for the $T_C(d)$ data set of Zhao \textit{et al.}\cite{Zhao} for films also on sapphire. We conclude that the obtained thickness of NbO$_x$ in our case ($\sim 1$ nm) is reasonable, being comparable but larger than found in films grown on sapphire. This analysis confirms the important role played by the substrate. In particular, the absence of reactive oxygen, especially onto the substrate surface, is a mandatory requirement to deposit ultra-thin and ultra-clean Nb films, avoiding or drastically limiting the formation of a metallic oxide at the film-substrate interface.

\begin{figure}[ht]
\centering
\includegraphics[width=\linewidth]{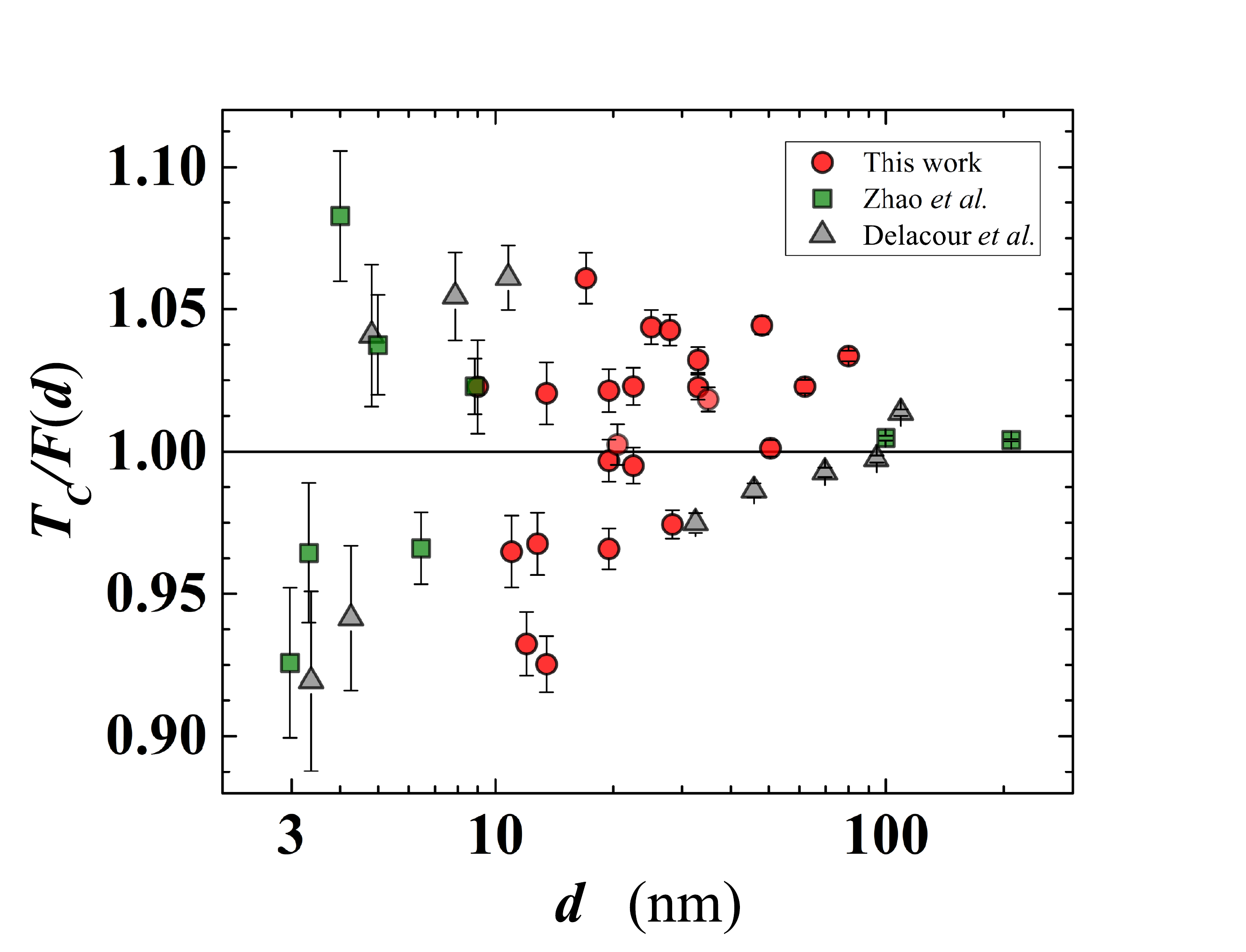}
\caption{Plot of the ratio $T_C/F(d)$ (see text) as a function of the film thickness for several Nb data sets. The fitting function, $F(d)$, has been calculated using the theoretically expected value of the Nb bulk transition temperature (i.e. $T_{C0} = 9.22$ K). Circles: present work; triangles: Zhao \textit{et al.}\cite{Zhao}; squares: Delacour \textit{et al.}\cite{Delacour} The error bars are defined by Eq. (S7) of the supplementary information.}  
\label{Fit_ratio_Tc_vs_d}
\end{figure}

At this point it is worth noting that in the presence of an overall suppression of $T_C(d)$ due to proximity effect, the fitting function given by Eq. \eqref{Tcantiprox} can be subtracted from the general data set in order to amplify the visibility of the remaining oscillations of $T_C$ for decreasing $d$. Namely, in this range of thicknesses, one expects that $T_C$ starts to display oscillations due to proliferating quantum size effects and shape resonances associated with confinement effects in the perpendicular direction. We have applied the aforementioned subtraction to our $T_C(d)$ data, but also to the data of Delacour \textit{et al.} \cite{Delacour} and Zhao \textit{et al.}\cite{Zhao} where such oscillations were not considered. Remarkably, in all cases we detected residual $T_C$ oscillations of increasing amplitude for decreasing $d$ (shown in Figure \ref{Fit_ratio_Tc_vs_d}). The observed increase of oscillations could be attributed to an incipient effect of the shape resonances when $d < 20$ nm. Namely, the detected oscillations of $T_C$ with amplitude of ~5\% in the thinnest films are very much comparable to the theoretical predictions for Al and Pb films of similar thickness, discussed in Ref.\cite{Shanenko2006B} (and references therein), suggesting their quantum-confinement origin in our Nb nanofilms. As should be the case, for larger thicknesses ($d >20$ nm) the $T_C$ oscillations in our films get progressively reduced in amplitude.

\begin{figure}[ht]
\centering
\includegraphics[width=\linewidth]{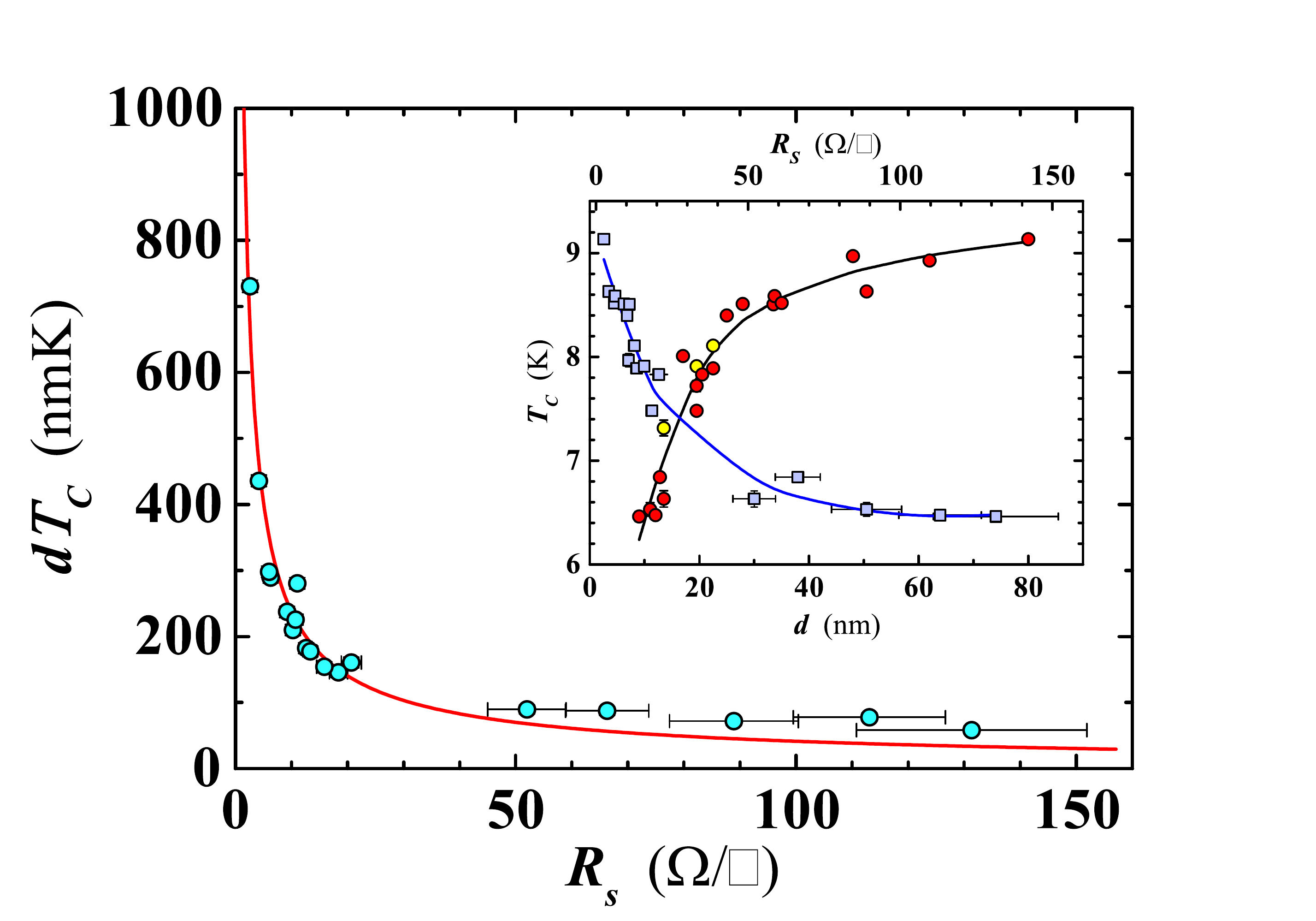}
\caption{Dependence of the product of the film thickness, $d$, and the superconducting transition temperature, $T_C$, on the sheet resistance ($R_S$) for the investigated Nb nanofilms. The red line is the least-squares fit of the data obtained using Eq. \eqref{eq1}. Inset reproduces the behavior of $T_C$ (circles) and of $R_S$ (diamonds) as a function of the thickness. Yellow filled symbols refer to samples deposited on sapphire substrates. Blue and black lines are guides for the eyes.}  
\label{FigdTc_vsRs}
\end{figure}

As the final characterization of the critical temperature in our films, we discuss the recent findings of Ivry \textit{et al.}\cite{Ivry2014} that $T_C$ should exhibit an intricate dependence not only on the film thickness but also on the sheet resistance, $R_S$. Namely, Ivry \textit{et al.} demonstrated a universal relationship among $T_C$, $d$ and $R_S$, for films with thicknesses in a broad range (i.e. $\approx 1 \div 10^3$ nm) and belonging to different classes of superconducting materials, \cite{Ivry2014} showing that experimental data can be well described by the relation:
\begin{equation}
\centering
\label{eq1}
d\cdot T_C = A{R_S}^{-B},
\end{equation}
where $A$ and $B$ are fitting parameters, hereafter considered unitless. We have therefore investigated the behavior of $T_C$ in our films as a function of $R_S$, and tested the validity of Eq. \eqref{eq1} in our case. Figure \ref{FigdTc_vsRs} shows the very successful fit, yielding $A = 1350 \pm 120$ and $B = 0.76\pm 0.05$. Our $B$ value is in excellent agreement with the result of Ivry \textit{et al.} for Nb films, while our parameter $A$ is larger.\cite{Ivry2014,Ivry2014_Supplemental} This fact, in addition to the better quality of our nanofilms, is also due to the different range of data points used for the fit compared to those considered in Ref.\cite{Ivry2014} The inset of Figure \ref{FigdTc_vsRs} shows that in our case $T_C$ steeply decreases for $0 < R_S \lesssim 20$ $\Omega/\square$, and then attains an almost constant value for $R_S > 50$ $\Omega/\square$. 

\subsection*{Other critical quantities and characteristic length scales of the superconducting state}

\begin{figure}[ht]
\centering
\includegraphics[width=\linewidth]{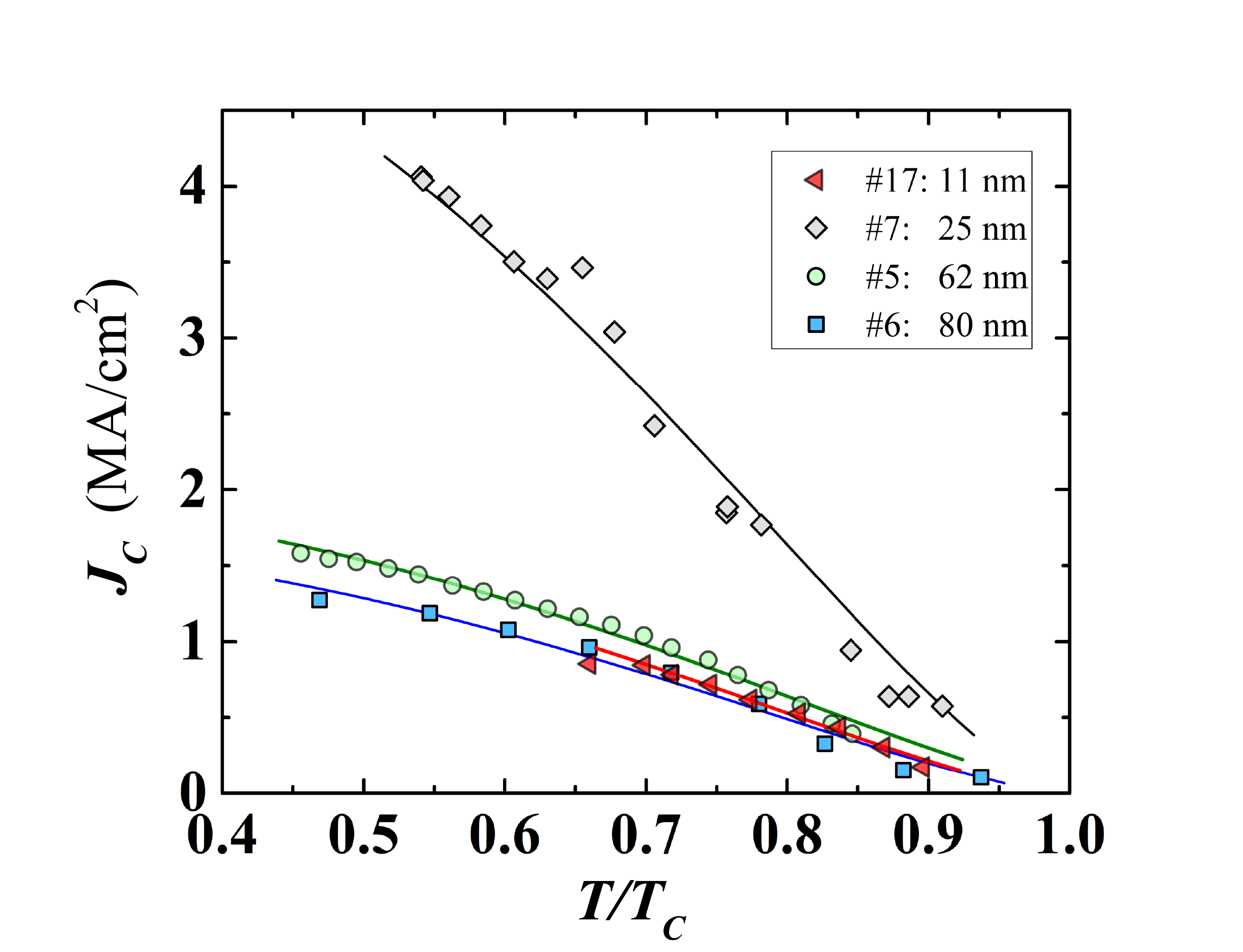}
\caption{Normalised temperature dependence of the critical current density. Plotted curves refer to four films covering the whole range of thicknesses of the considered Nb nanofilms. Lines are the least-squares fit by the Ginzburg-Landau equation S1 (see the supplementary material) to the experimental data points.}
\label{Jc_vs_d}
\end{figure}

Additional information about the physical properties exhibited by the here investigated Nb films can be extracted from the behavior of the critical current density, $J_C$, and the maximal magnetic field the superconducting samples can sustain, i.e. the upper critical magnetic field $H_{C2}$.
We realised measurements of the critical current in absence of any applied magnetic field, determined as a current corresponding to the onset of normal-state resistance in the current-voltage curves for the investigated films, at different fixed temperatures. The aim is to obtain the functional temperature dependence of these quantities and then extrapolate their values at zero temperature. The dependence of the critical current density, $J_C$, on reduced temperature, $t = T/T_C$, for selected films (representative of all investigated ones), is shown in Figure \ref{Jc_vs_d}. 
To model the measured $J_C(t)$ behaviour, we rely on the Ginzburg-Landau (GL) theory, although it is formally valid only near $T_C$. However, to recover the experimentally well-known temperature dependence of thermodynamic critical field $H_c\propto(1-t^2)$ (as opposed to $H_c\propto(1-t)$ in standard GL theory), we employ the empirical modifications proposed by Ginzburg\cite{Ginzburg1956} (also corresponding to the temperature dependence from a two-fluid model), as already successfully used in theoretical descriptions of single-crystalline superconductors even far below $T_C$.\cite{Muller2012} As a consequence of these modifications, the GL parameter $\kappa=\lambda/\xi$ (where $\lambda$ is the penetration depth and $\xi$ the coherence length) becomes temperature dependent $\kappa(T)=\kappa(0)\big/(1+t^2)$, the upper critical field exhibits dependence $H_{c2}\propto(1-t^2)\big/(1+t^2)$, and the depairing current density becomes proportional to $\left(1-t^2\right)\sqrt{1-t^4}$. For more details, we refer the reader to supplementary material.

\begin{figure}[ht]
\centering
\includegraphics[width=\linewidth]{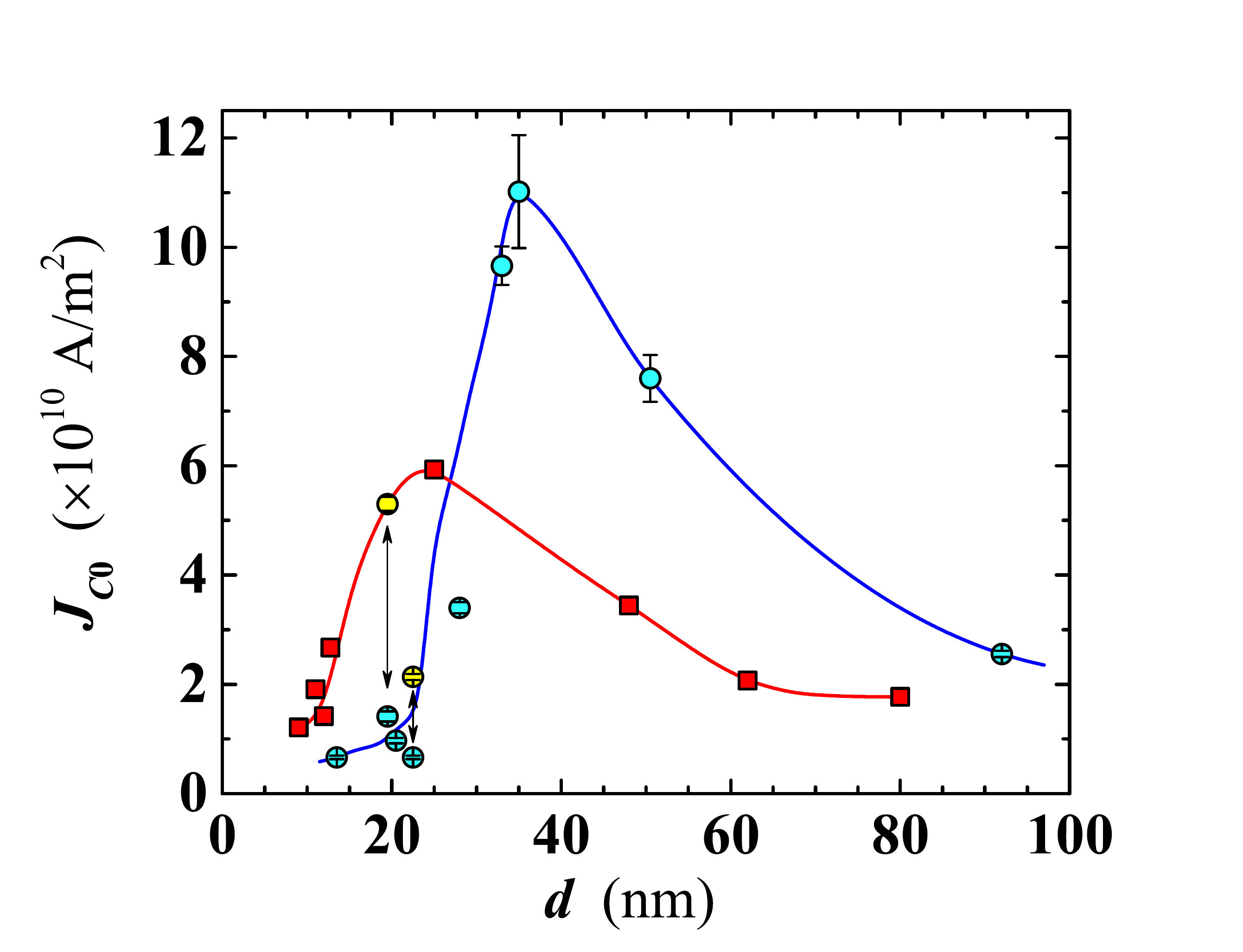}
\caption{Thickness dependence of the critical current density. Circles: films $w = 10$ $\mu$m wide; yellow filled symbols correspond to the films deposited on sapphire substrates. Squares: $w = 50$ $\mu$m wide. The two lines are guides for the eyes.}  
\label{Jc0_vs_t}
\end{figure}

In absence of applied magnetic field, the measured critical current density should (nearly) correspond to the depairing current density. Assuming the above modified GL behavior of $J_C(T)$, we performed the fitting of the obtained data to extract $J_{C0}$ for films of different thicknesses, having width of either $w =10$ $\mu$m or $w =50$ $\mu$m. We found a strong dependence of $J_{C0}$ on both the thickness and the width. It turns out that, compared to the Nb bulk $J_{C0}$ value of 1.96 MA/cm$^2$, our films exhibit considerably higher $J_{C0}$ for thickness in the range $13 < d < 60$ nm, with a peak value $\sim 3$ and $\sim 5.5$ times higher for sample width $w =50$ $\mu$m and $w =10$ $\mu$m, respectively (see Figure \ref{Jc0_vs_t}). The sizeable increase of $J_{C0}$ upon narrowing the film is related to an enhancement of the pinning effect of the vortices as observed also by Il'in \textit{et al.} \cite{Ilin2005} In addition, the increase of $J_{C0}$ in nanofilms deposited on sapphire evidences once more the important role played by this substrate in improving the superconducting properties of the nanofilms.

\begin{figure}[ht]
\centering
\includegraphics[width=\linewidth]{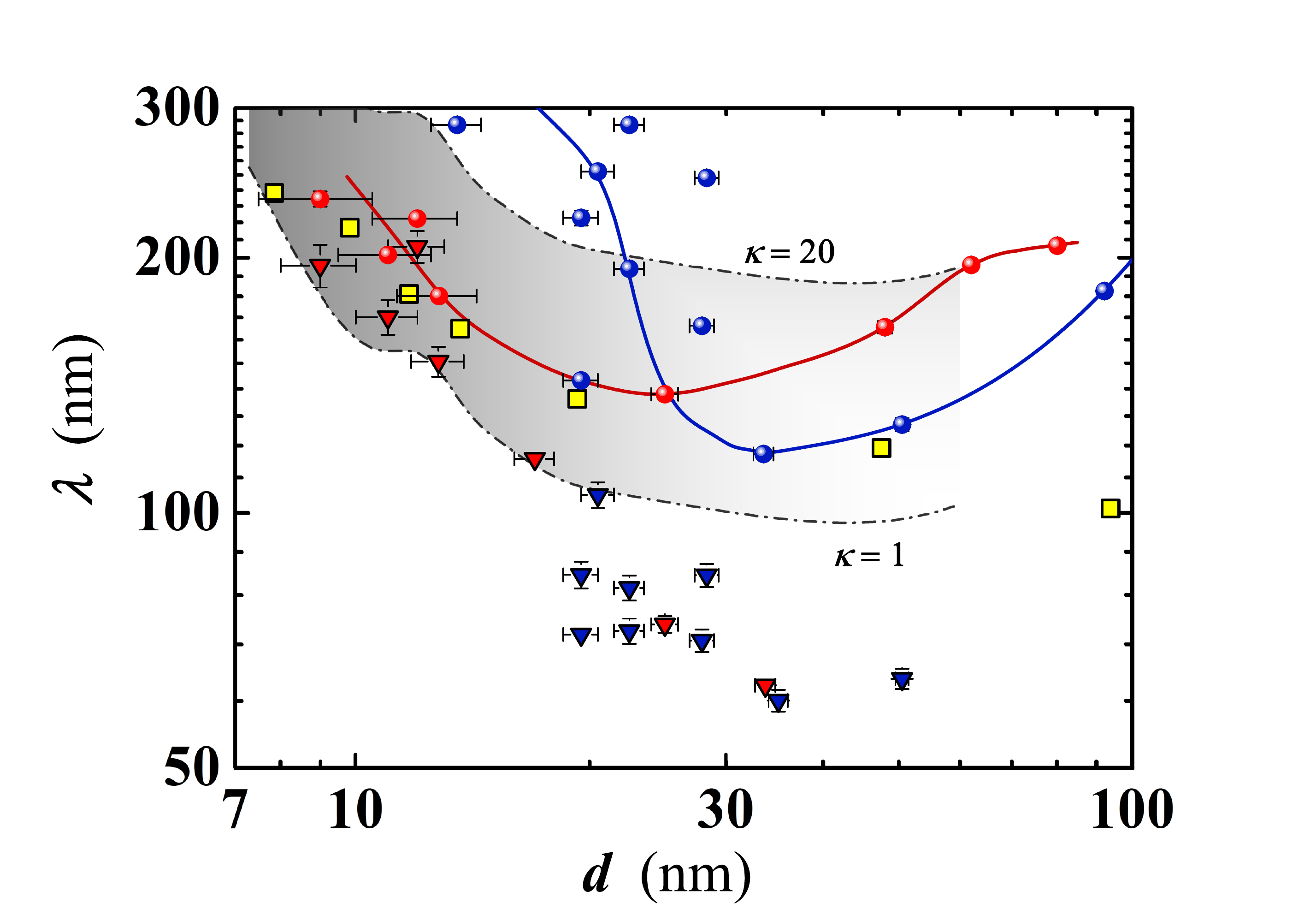}
\caption{Penetration depth as a function of the Nb film thickness. Values have been calculated by using both Eq. \eqref{eqTT}, with $\kappa = 2.2$ (spheres), and Eq. \eqref{Lamb} valid in the dirty limit (triangles). Shown data refer to films having width $w =10$ $\mu$m (blue) or $w = 50$ $\mu$m (red). Squares: values obtained by Gubin \textit{et al.}\cite{Gubin2005} Dash and dotted lines delimit the range of $\lambda$ variation (for the case $w = 50$ $\mu$m), for $\kappa$ varied between 1 (bottom line) and 20 (top line) in Eq. \eqref{eqTT}. Continuous lines are guides for the eyes.}  
\label{lambda}
\end{figure}

According to the criterion recently proposed by Talantsev and Tallon\cite{Talantsev2015}, we apply the relation between the critical current density and the London penetration depth $\lambda$, for type-II superconductors with thickness smaller than $\lambda$ ($2d<\lambda$):
\begin{equation}
\centering
J^{II}_C(sf) = \frac{H_{C1}}{\lambda}=\frac{\Phi_0}{4\pi\mu_0\lambda^3}\left(\ln\kappa + 0.5\right),
\label{eqTT}
\end{equation}
where $\Phi_0$ is the magnetic flux quantum; $\mu_0$ the vacuum permeability and $\kappa = \lambda / \xi$ generally assumed to be $\approx 1$ for Nb in the clean limit. $J^{II}_C(sf)$ represents the current density which generates the self field that induces vortices and itself limits the maximal current in absence of any applied magnetic field.
Applying Eq. \eqref{eqTT} to the measured values of the critical current density we have extrapolated the values of $\lambda$ at the lowest accessible $T$ by the experimental setup ($\approx 4$ K). The resulting thickness dependence of $\lambda$ is shown in Figure \ref{lambda} together with the $\lambda$ values measured in Nb films by Gubin \textit{et al.}, using a resonance technique.\cite{Gubin2005} The agreement between our data and those of Ref. \cite{Gubin2005} is very good up to $d \approx 25$ nm, with an increasing discrepancy for $d \gtrsim 50$ nm. 
An alternative check of the thickness dependence of $\lambda$ has been done by using the experimental mean free path $l$ (Figure \ref{Lmean}) and the relation valid in the dirty limit:\cite{deGennes}
\begin{equation}
\centering
\lambda = 0.62\lambda_L\sqrt{\left(\frac{\xi_0}{l}\right)},
\label{Lamb}
\end{equation}
where $\lambda_L = 39$ nm and $\xi_0 = 38$ nm are the Nb bulk value of the London penetration depth and the BCS coherence length, respectively. For thinner films ($d \lesssim 20$ nm) the values of $\lambda$ calculated either by Eq. \eqref{Lamb} or Eq. \eqref{eqTT} are comparable while diverging, in values and behaviour for thicker films. 
While $\lambda$ calculated by Eq. \eqref{Lamb} preserves the likely correct $d$ dependence and decreases towards the bulk limit, its values are more than halved with respect to those obtained in Ref. \cite{Gubin2005} The apparent disagreement in our $\lambda(d)$ data calculated by the two methods, can be accounted for considering that Eq. \eqref{eqTT} is valid for $2d \ll \lambda$,\cite{Talantsev2015} condition that appears to be fulfilled in our case only below $d \approx 20$ nm (Figure \ref{lambda}). Finally, it is worthwhile to consider also the dependence of $\lambda$ on the value of $\kappa$ in Eq. \eqref{eqTT}, considering that much larger values of the GL parameter are expected in dirty films compared to single-crystalline samples.
Figure \ref{lambda} reproduces also the possible range of values taken by $\lambda$ for $1 \leq \kappa \leq 20$ in the thickness range $10 < d < 50$ nm. Looking at Figure \ref{lambda} it is evident that assuming $\kappa \approx 1$ (as was actually done in Ref.\cite{Talantsev2015}) a better agreement between the two calculation methods of $\lambda$ would have been found.

\begin{figure}[ht]
\centering
\includegraphics[width=\linewidth]{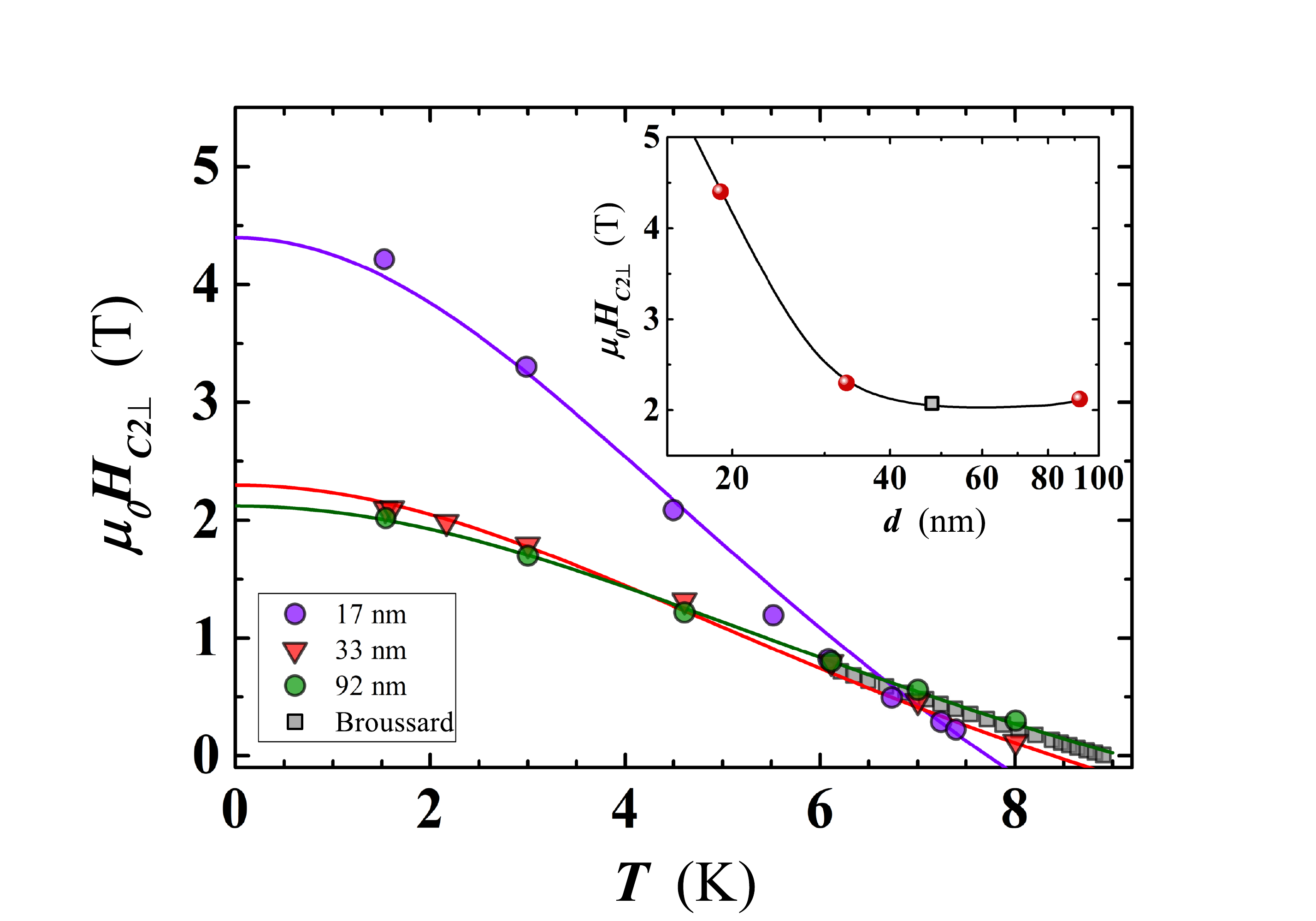}
\caption{Temperature dependence of the perpendicular critical magnetic field ($H_{C2\perp}$). For each temperature, the corresponding $H_{C2\perp}$ value has been derived as the mean value of the applied magnetic fields corresponding to $10\%$ and $90\%$ of the resistivity value at the saturation point (see Figure S3 in supplementary material). Lines are the least-squares fit of the data points by Eq. \eqref{HC0}.
For comparison, data of Broussard\cite{Broussard} for the $H_{C2\perp}$ of a 50 nm thick Nb film have been added to the plot.
Inset: thickness dependence of $H_{C2\perp}$ at 1.6 K, for the three investigated Nb films (spheres) and for the sample of Broussard\cite{Broussard} (square) whose value, at 1.6 K, has been extrapolated by Eq. \eqref{HC0} from the experimental data available down to $\approx 6$ K. The black line serve as a guide for the eyes.}  
\label{Hc2}
\end{figure}

Finally, we have also studied the temperature dependence of the perpendicular critical magnetic field, $H_{C2\perp}$, for three Nb nanofilms having a thickness of 19.5 nm, 33 nm and 92 nm ($\#H1$, $\#H8$ and $\#H12$, respectively). The $H_{C2\perp}$ value has been determined by measuring the resistivity as a function of the applied magnetic field, at different fixed $T$ (Figure S3 in supplementary material).
The experimental $H_{C2\perp}(T)$ curves\footnote{At a fixed $T$, the $H_{C2\perp}(T)$ has been determined as the average of the two $H$ values corresponding to the $10\%$ and the $90\%$ of the $\rho(H)$ saturation value.} for the three films are reproduced in Figure \ref{Hc2}. The two thicker films (i.e. $\#H8$ \& $\#H12$) exhibit practically overlapped curves. Approaching the lowest $T$, $H_{C2\perp}(T)$ tends to saturate while a small deviation from the linearity has been detected for the thinner film (i.e. $\#H1$) at $T > 5.5$ K, suggesting inhomogeneity effects in the film thickness and/or presence of defects due to the reduced $d$ value, that is likely related to an interaction of the Nb film with the SiO$_2$ of the substrate, as mentioned above in the discussion of the rapid decrease of $T_C$ in that limit. 
The $T$ dependence of $H_{C2\perp}$ has been captured by a least-squares fit of the experimental values using the relation:\cite{Bao2005}
\begin{equation}
\centering
H_{C2\perp}(T) = H_{C20\perp}\frac{\left[1-\left(T/T_C\right)^2\right]}{\left[1+\left(T/T_C\right)^2\right]},
\label{HC0}
\end{equation}
with $H_{C20\perp}$ being the value of the orthogonal upper critical magnetic field at zero temperature.

The values of $H_{C2\perp}(T)$ allow us to calculate the in-plane coherence length $\xi(T)$ of our films, extracted by the standard GL relation:\cite{Bao2005}
\begin{equation}
\centering
H_{C2\perp}(T) = \frac{\Phi_0}{2\pi\xi^2(T)},
\label{eq3}
\end{equation}
The Ginzburg-Landau coherence lengths at $T = 0$ K, $\xi(0)$, for the three samples considered in Figure \ref{Hc2}, have been derived using the Eq. \eqref{eq3} and the $H_{C20\perp}$ values extracted by the Eq. \eqref{HC0}. $\xi(0)$ raises from $\simeq 7.8$ nm for $d = 19.5$ nm to $\approx 11.5$ nm for $d \geq 90$ nm, compared to the Nb bulk value of $\xi(0) = 0.74\xi_0\approx28$ nm.

Our results are in agreement with those reported by Trezza \textit{et al.} for Nb films of comparable thickness,\cite{Trezza} and with values measured by Broussard in a 48 nm thick Nb film.\cite{Broussard} Figure \ref{Hc2} reproduces the $H_{C2\perp}(T)$ data of Ref. \cite{Broussard} and shows an overlapping trend with data for our two thicker films, suggesting a saturation of $H_{C2\perp}$ for Nb films with $d$ larger than 33 nm. The hypothesis has been confirmed also by the $H_{C2\perp}(T)$ value of the Broussard film at 1.6 K obtained through a best fit with Eq. \eqref{HC0} (see the inset of Figure \ref{Hc2}). 

\begin{figure}[h]
\centering
\includegraphics[width=\linewidth]{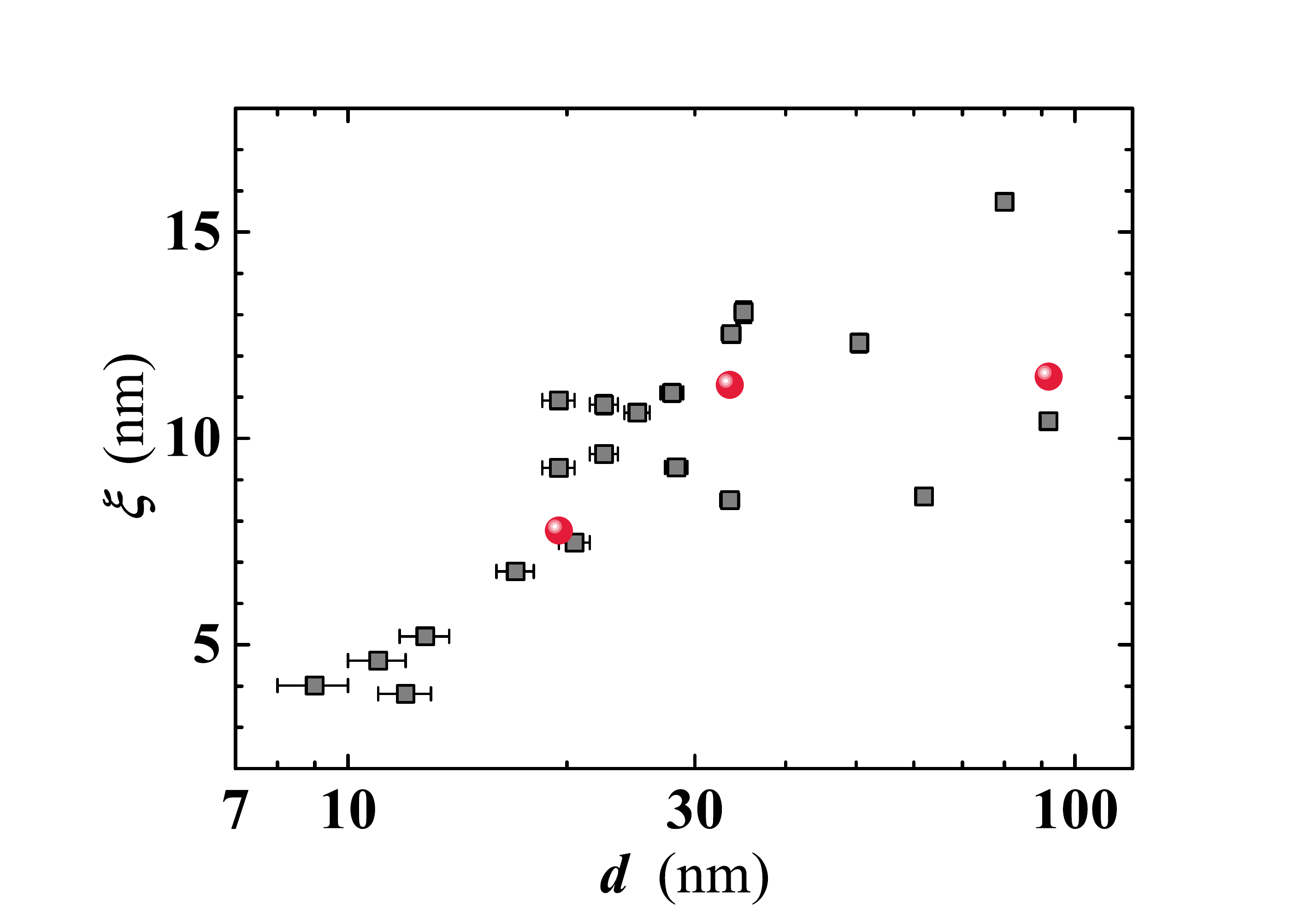}
\caption{Thickness dependence of the coherence length at zero temperature for investigated Nb films. Squares: values derived by Eq. \eqref{xi_dirty} taking into account the measured mean free path ($l$). Spheres are the $\xi_0$ values calculated by the measured $H_{C2\perp}(T)$ and the fitting procedure described in the text.} 
\label{xi}
\end{figure}

Finally, the thickness dependence of the superconducting coherence length in our samples has been evaluated using the experimental $l$ values, under the validity of the dirty limit condition:
\begin{equation}
\centering
\xi(0) = 0.855\sqrt{\left(\xi_0 l\right)},
\label{xi_dirty}
\end{equation}
where $\xi_0 = 38$ nm is the BCS Nb bulk coherence length. The value of the coherence length at 10 K raises from $\xi = 4$ nm for $d \approx 10$ nm, up to $\xi \simeq 12$ nm for $d \approx 35$ nm. For larger thicknesses we observe a saturation effect albeit data points appear somewhat scattered (Figure \ref{xi}). In any case, the $\xi$ values calculated by Eq. \eqref{xi_dirty} are in very good agreement with those derived from the $H_{C2\perp}(T)$ measurements (Figure \ref{xi}). Finally we note that, for the nanofilm with thickness of 19.5 nm, $\xi(0) \simeq 7.8$ nm evidences presence of smaller Cooper pairs, exhibiting shrinking by a factor 5 compared to the bulk value. 

\section*{Discussion and Conclusions}
The comprehensive experimental characterisation of metallic and superconducting properties of Nb nanofilms reported in this work demonstrates two different regimes depending on the layer thickness. For large thickness, in the range 25 nm $< d < 80$ nm, all measured electrical properties evolve smoothly for decreasing $d$. The Nb nanofilms are found to be of high quality and superconductivity has a 3D character. Around $d = 25$ nm normal and superconducting state properties of the Nb nanofilm display an abrupt change in their behavior. Below 25 nm of thickness, superconductivity becomes progressively 2D, realizing a 3D-2D dimensional crossover. 
This dimensional crossover for decreasing thickness results to be particularly evident in the sudden suppression of $T_C$, in the peaked behaviour of the critical current density and when comparing the evolution of the penetration depth and the correlation length as a function of thickness.
Approaching $d = 10$ nm, novel quantum phenomena not yet observed in Nb nanofilms start to emerge: (i) on top of the overall $T_C$ drop for decreasing $d$, remaining oscillations of $T_C(d)$ indicate incipient quantum size effects.  
(ii) A factor five enhancement of the penetration depth for the thinnest films with respect to the bulk has been observed, with a consequent strengthening of the type-II magnetic behavior of the nanofilms with thickness below 20 nm. The magnetic penetration depth has been extracted from the critical current measured in this work at the lowest temperature $T\simeq 4$ K.
Note that the here reported data for the penetration depth $\lambda(d)$ compare in a satisfactory way with other measurements of $\lambda$ realized in the range of low thickness $d \leq 25$ nm. 
(iii) A sizeable amplification of the upper critical magnetic field is associated with a considerable shrinking of the Cooper pair size, locating the thinnest Nb nanofilms close to the BCS-BEC crossover regime, predicted for very thin nanofilms.
We note that close to $d = 25$ nm the Nb nanofilms show optimized superconducting properties: maximal $J_C \simeq 6$ MA/cm$^2$, $\lambda = 150$ nm and $\xi(0)\simeq 10$ nm, $H_{C2\perp}(T) = 2$ T, $T_C = 8$ K, and a very narrow superconducting transition of $\Delta T_C = 40$ mK. Therefore, Nb nanofilms with $d\simeq 25$ nm on an insulating substrate (here we considered SiO$_2$ and Al$_2$O$_3$) constitute an ideal platform for further nanostructuring in the form of single or periodic stripes or pattern of dots and for the realization of nano superconducting devices.
In addition, we have shown that the thickness dependence of the mean free path of the carriers plays a crucial role in understanding the thickness dependence of the superconducting properties, mainly in the penetration depth and in the coherence length. The effects of disorder are therefore entangled with the observed dimensional crossover and with the BCS-BEC crossover occurring below few tens of nanometers in our Nb nanofilms.
We also found that the amplification effects of the superconducting properties reported in this work depend on the substrate and on the lateral width of the nanofilms. Optimal superconducting properties have been demonstrated using sapphire as a substrate and squeezing the lateral size of the nanofilms toward few micrometers, furthermore suggesting great potential in the fabrication of superconducting nanostripes of Nb at the nanometer scale.

\section*{Methods}

Niobium nanofilms have been deposited at room temperature on thermally oxidized Si wafer (silicon oxide thickness: $300\div500$ nm), by an ultra vacuum DC sputtering system in a base pressure of about $2\times 10^{-9}$ mbar. Film thickness has been varied from about 9 nm to 90 nm, keeping constant the deposition rate at 0.65 nm/s (see Table I). Scanning electron microscopy (SEM) analysis has been carried out on some films by a FEI Quanta$^{TM}$ 3D FIB (Nanofacility Piemonte, INRIM).

For characterization of the electrical properties samples have been shaped in a Hall bar geometry, $1 \div 2$ cm long, 10 $\mu$m and 50 $\mu$m wide.
The resistivity, $\rho(T)$, and the current-voltage (I-V) characteristics have been measured as a function of the temperature, in the range 4 $\div$ 300 K, by a He closed cycle cryostat (Advanced Research System mod. 210 DE) equipped with two silicon diode thermometers (Lakeshore DT-670, one of which was calibrated with a maximum error of 6.3 mK) and a temperature controller Lakeshore mod. 332. Resistivity and I-V characteristics have been measured sourcing a constant current (Keithley mod. 220), monitored either by a pico-ammeter (Keithley mod. 6487 for $\rho(T)$) or by a multimeter (Hewlett-Packard mod. 34401A for I-V). The voltage drop has been detected by a multimeter (Keithley mod. 2000). For the measure of $\rho(T)$ the current sourced has been in the range $1 \div 50$ $\mu$A.
Depending on the kind of characterization, measures have been executed either with (e.g. I-V characteristics) or without $T$ stabilization (e.g. $\rho(T)$). In the former case, a $T$ stability better than 50 mK has been achieved below 15 K. In the latter approach, the variation of $T$ during data acquisition was lower than $20\div30$ mK. 
The superconducting transition temperature $T_C$ as well as the width of the transition was measured by a liquid He-cryostat equipped with a silicon calibrated thermometer. 
For the determination of the upper critical magnetic field, $H_{C2}$, a liquid $^4$He cryostat, Oxford Instruments Teslatron 16T has been used, equipped with a superconducting magnet (up to 16 T) and a variable temperature insert working from 300 K down to 1.5 K. The resistance values were measured by using either a PICOWATT AVS-47 AC or a Lakeshore mod. 370 AC resistance bridge.

\section*{Acknowledgements}
We thank Antonio Bianconi, Mauro Doria and Vincenzo Lacquaniti for useful discussions. We acknowledge the collaboration with Federica Celegato for AFM analysis and Sara Quercetti for the electrical properties characterization.
A.P. and N.P. acknowledge financial support from University of Camerino FAR project CESEMN. We also acknowledge the collaboration within the MultiSuper International Network (http://www.multisuper.org) for exchange of ideas and suggestions.

\section*{Author contributions statement}
N.P., A.P. and N.DL. conceived the experiments. M.F. and N.DL. fabricated samples. N.P., S.J.R., M.F. and C.C. conducted the experiments, N.P., A.P., L.F. and M.V.M. analysed the results. All authors reviewed the manuscript.

\section*{Additional information}
Supplementary information accompanies this paper. The authors declare no competing financial interests.

\section*{\Large{Supplementary Information}}

\section{Methods}

\subsection{Experiment}
\begin{figure}[ht]
\renewcommand{\thefigure}{S\arabic{figure}}

\centering
\includegraphics[scale=0.55]{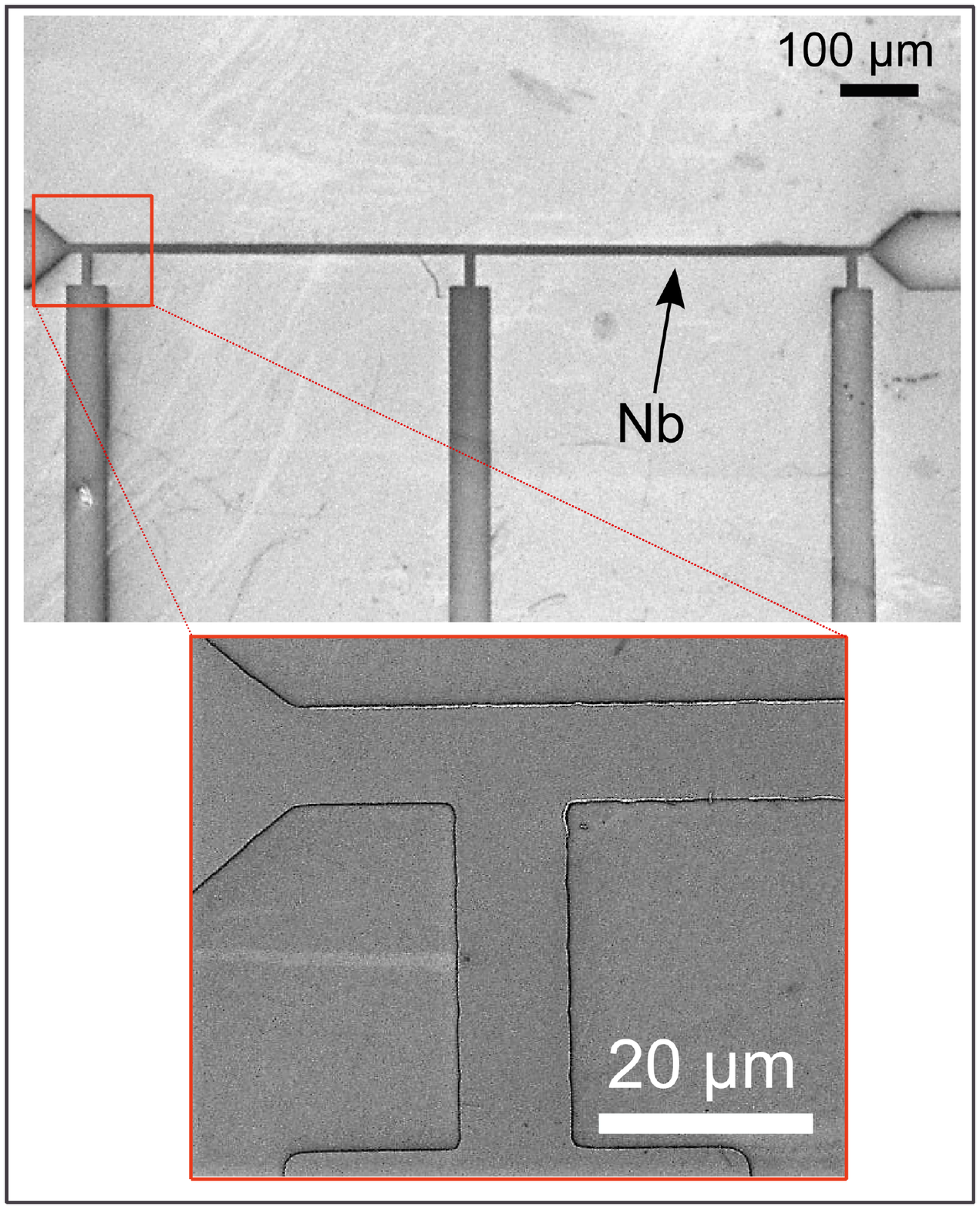}
\caption{Scanning electron microscopy of a Hall bar device obtained by a 25 nm thick Nb film deposited on SiO$_2$. Five large Nb contacts, two horizontal (at the midlle of the image) and three vertical are used to source the current and to measure the voltage drop (a couple of them), respectively. Inset: further magnification of the contact area zone (red box) showing a uniform and homogenous formation of a Nb layer also at the contact area.}
\label{SEM}
\end{figure}

For electrical and magnetic characterization of the Nb films properties, a typical Hall bar geometry has been fabricated. Niobium has been used as material for the fabrication of both pads and lateral arms (for the voltage probe) (Figure \ref{SEM}). 

\subsection{Superconducting critical current density}
\begin{figure}[ht]
\renewcommand{\thefigure}{S\arabic{figure}}
\centering
\includegraphics[scale=0.50]{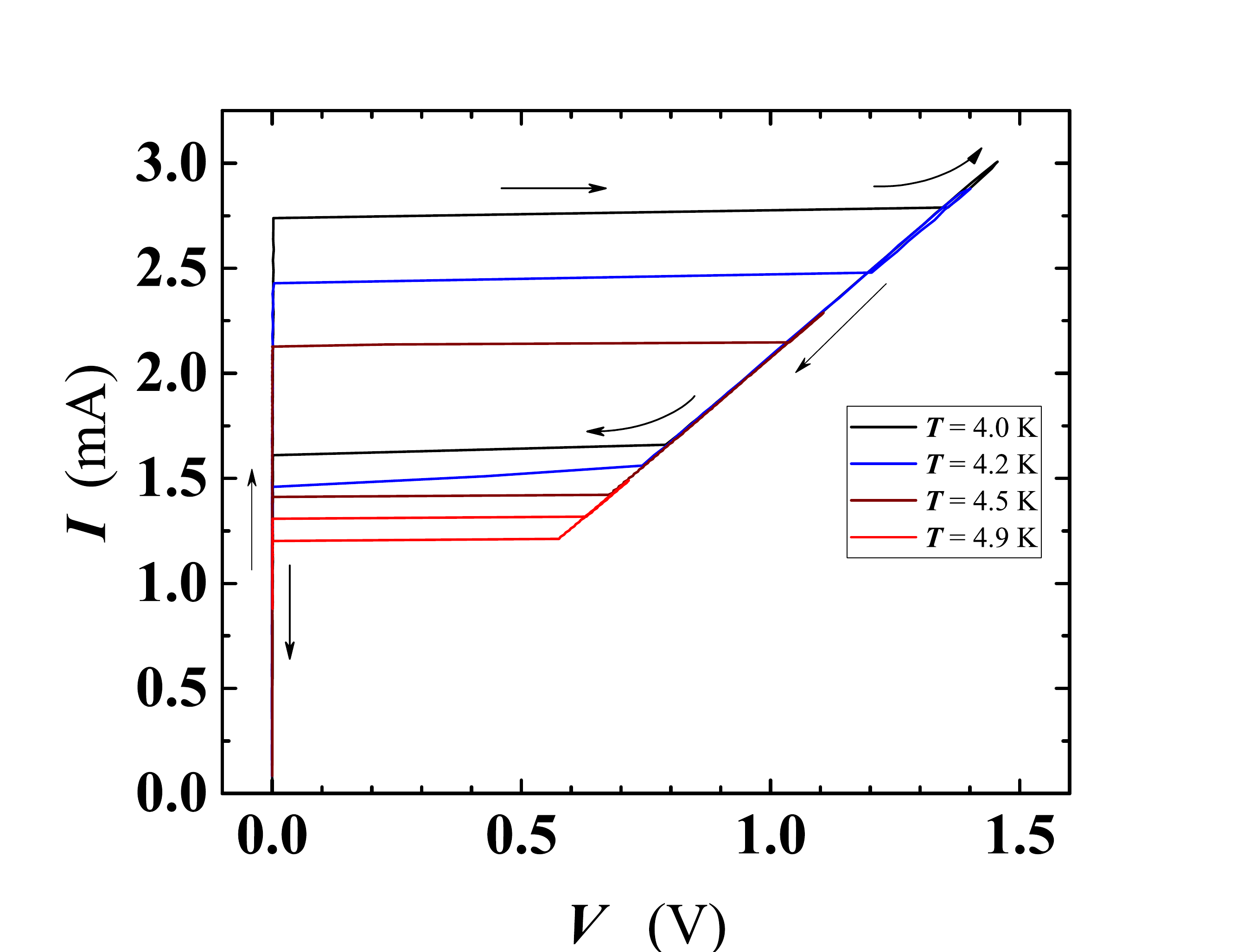}
\caption{Typical hysteretic current-voltage curves, at few selected $T$, for the film $\#7$ (see the Table I), used to calculate the superconducting critical current density (see Figure \ref{Jc0_vs_T}).  The arrows indicate the voltage sweep direction for the curve measured at 4 K.}  
\label{I-V}
\end{figure}

The value of $J_{C0}$ has been derived by the current-voltage characteristic carried out sourcing slow-sweep currents at fixed $T$, in the range from 4 K to $T_C$. Hysteretic I-V curves  have been measured for the investigated films, similar to that shown in Figure \ref{I-V}. We assumed as critical current density, $J_C$, the $J$ value corresponding to the jump into the normal state in the I-V curve (Figure \ref{I-V}). The superconducting critical current density at zero $T$, $J_{C0}$ has been derived by a least squares fit of the $J_C(T)$ curve
assuming the mean-field behavior of $J_C(T)$ predicted by the Ginzburg-Landau (GL) theory for depairing critical current density:
\begin{equation}\tag{S1}
\label{eqJc}
J_C(T) = J_{C0}\left[1-(T/T_C)^2\right]^{3/2}\left[1+(T/T_C)^2\right]^{1/2}.
\end{equation}
We experimentally check this $T$ dependence of the $J_C(T)$ (Eq. \ref{eqJc}) for several Nb layers covering the range of thicknesses and the two Hall bar widths (10 $\mu$m and 50 $\mu$m) used in the present study. 

\subsection{Error analysis of $T_c$ oscillations}

According to the McMillan model, the $T_C$ suppression law is related to the thickness $d$ by the following relation 
\begin{equation}\tag{S2}
\label{Mcm}
T_C(d) = T_{C0}\left(\frac{3.56\times T_D}{\pi\times T_{C0}}\right)^{-\frac{\alpha}{d}} = T_{C0}\,\exp\biggl\lbrace{-\ln\left(\frac{3.56\times T_D}{\pi\times T_{C0}}\right)\frac{\alpha}{d}\biggr\rbrace},
\end{equation}
where $T_D = 277$ K is the Debye temperature, and $T_ {C0} = 9.22$ K is the Nb bulk critical temperature. The quantity $\alpha$ is the parameter of the model which corresponds to the normal layer thickness $d_N$ when we set the superconducting-to-normal density ratio equal to unity.

The method of least squares is an analytical technique for finding the fitting curve of the model which optimally describes a set of data. Using the McMillan model, the measured values of $d$ and $T_C$ are inserted into the Eq.\ref{Mcm} to determine the optimal parameter $\alpha$ which defines the behaviour of the nonlinear fitting curve $F(d)$. For semplicity we use unit weighting for all data points (from $1$ to $n$) because we cannot infer any estimate of their uncertainties and therefore we assume that every datum point has the same weight. In order to get the parameter $\alpha$, the procedure for the least squares solution needs to be iterative. For each iteration the solution is obtained with the method described in ref.\cite{Wol} 

Accordingly, here we report the procedure related to the least squares solution for the McMillan model. Starting with an initial guess of $\alpha_\text{init}=1$ and with the set of data $(d_i,T_C^{\,i})$ ($\forall i \in \lbrace{1,\ldots,n}\rbrace$), we define

\begin{equation}\tag{S3}
\label{mat_elem}
C=\sum_{i=1}^n \left[g^i_{\alpha}\right]^2 \qquad V=\sum_{i=1}^n T_C^{\,i}g^i_\alpha,
\end{equation}
where $g^i_\alpha=\frac{\partial T_c(d_i)}{\partial\alpha}$. It is worth noting that in the equation for $V$ we consider values of the dependent variable $T_C^{\,i}$ minus the computed value $T_C(d_i)$ obtained with $\alpha_\text{init}$. 

The parameter $\alpha_\text{calc}$ is calculated by solving the matrix equation
\begin{equation}\tag{S4}
\label{Avect}
\alpha_\text{calc}=C^{-1}V
\end{equation}
where calculated value of $\alpha_\text{calc}$ will no longer be the final solution but it represents the change in the value of the initial guess $a_\text{init}$
\begin{equation}\tag{S5}
\alpha_\text{new}=\alpha_\text{init} +\alpha_\text{calc}
\end{equation}
and the value of $\alpha_\text{new}$ is then used as initial guess $\alpha_\text{init}$ for the next iteration. This process is continued until convergence, i.e., when the fractional change is less than a specified value of $\varepsilon$ (we use $\varepsilon = 10^{-5}$). 
\begin{equation}\tag{S6}
\label{conv_crit}
\bigg|\frac{\alpha_\text{calc}}{\alpha_\text{init}}\bigg|\leqslant\varepsilon
\end{equation}
where Eq.\ref{conv_crit} has to be modified if $\alpha_\text{init}$ is zero or very close to zero and then the condition will depend on the absolute value of $\alpha_\text{calc}$ and not on the relative value.

When the convergence is achieved and the optimal parameter $\alpha_\text{opt}$ is determined, the fitting curve $F(d)$ corresponds to the function $T_C(d)$ with $\alpha_\text{opt}$. The uncertainty $\sigma_F$ for every data point $i$ is:

\begin{equation}\tag{S7}
\sigma_F^{\,i}=\left(\frac{S}{n-1}\left[\frac{\partial F(d_i)}{\partial\alpha_\text{opt}}\right]^2 C^{-1}\right)^{1/2}
\end{equation}
where the weighted sum of squared residual $S=\sum_{i=1}^n\left[T_C^{\,i}-F(d_i)\right]^2$.

The ratio $\frac{T_C}{F(d)}$ is a good quantity for detecting the $T_C$ oscillations at low thicknesses. The uncertainty $\sigma\left(\frac{T_C}{F(d)}\right)$ is only determined by $\sigma_F$ and we get
\begin{equation}\tag{S8}
\begin{aligned}
\frac{T_C}{F(d)\pm\sigma_F}=\frac{T_C}{F(d)}\frac{1}{1\pm\frac{\sigma_F}{F(d)}}\simeq\frac{T_C}{F(d)}\left(1\pm\frac{\sigma_F}{F(d)}\right)=\frac{T_C}{F(d)}\pm T_C\frac{\sigma_F}{\left[F(d)\right]^2}
\end{aligned}
\end{equation}
where the approximation is due to the use of the infinite geometric series for $\|x\|=\|\pm\frac{\sigma_F}{F(d)}\|<1$ which is truncated at the second term.
The uncertainty $\sigma\left(\frac{T_c^{\,i}}{F(d_i)}\right)$ for every data point $i$ is therefore:
\begin{equation}\tag{S9}
\sigma\left(\frac{T_C^{\,i}}{F(d_i)}\right)=T_C^{\,i}\frac{\sigma_F^i}{[F(d_i)]^2}.
\end{equation}

\section*{Figures and Tables}

\begin{table}[h]
\centering 
\begin{tabular}{|l|l|l|l|l|l|l|l|}  %sono delle elle minuscole intervallate dalla barra verticale che serve per chiudere con delle linee verticali la tabella
\hline
 Sample & Thickness & width & $T_C$ & $\Delta T_C$ & {$\rho_{300}$} & {$RRR$} & {$\mu_0H_{C20\perp}$} \\
 { }  &  {nm}  &  {$\mu$m}  &  {K}  &  {mK}  &  {$\mu\Omega$cm}  &  {($\rho_{300}/\rho_{10}$)}  &  {nm}\\
\hline
 {\#9}   &   {9}  &  {50}  & {6.462}  &  {50}  &  {118.2}  &  {1.827}  &  {- }\\
  \hline
 {\#17}   &  {11}  & {50}  & {6.53}  &  {65}  &  {97.8}  &  {2.00}  &  {-}\\ 
\hline 
 {\#12}   &  {12}  &  {50}  & {6.475}  &  {30}  &  {135.7}  &  {1.893}  &  {-}\\
 \hline
  {\#16}  &  {13}  &  {50}  & {6.84}  &  {19}  &  {84.9}  &  {2.208}  &  {-}\\
 \hline
 {\#H15}   &   {13.5}  & {10}  & {6.631}  &  {79}  &  {70.19}  &  {1.889}  &  {-}\\
  \hline
 {\#H14S}   &   {13.5}  & {10}  & {7.313}  &  {75}  & {-}  &  {-}  &  {-}\\
 \hline
 {\#H1S}   &   {19.5}  & {10}  & {7.91}  &  {35}  &  {31.0}  &  {2.563}  &  {-}\\
 \hline
 {\#H1}   &   {19.5}  &  {10}  & {7.72}  &  {55}  &  {20.7}  &  {2.367}  &  {$4.4\pm0.10$}\\
   \hline
 {\#H5}   &   {19.5}  &  {10}  & {7.48}  &  {49}  &  {35.8}  &  {2.165}  &  {-}\\
    \hline
 {\#H2}   &   {20.5}  & {10}  & {7.830}  &  {42}  &  {42.3}  &  {2.272}  &  {-}\\
 \hline
 {\#H11S}   &   {22.5}  &  {10}  & {8.108}  &  {28}  &  {28.4}  &  {3.191}  &  {-}\\
  \hline
 {\#H9}   &   {22.5}  &  {10}  & {7.888}  &  {44}  &  {30.1}  &  {2.675}  &  {-}\\
 \hline
 {\#7}   &   {25}  &  {50}  &  {8.399}  & {15}  &  {25.6}  &  {2.774}  &  {-}\\
  \hline
 {\#H13}   &  {28}   &  {10}  &  {8.51}   & {28} &  {25.8}  &  {3.049}  &  { }\\
 \hline
 {\#H7}   &  {28}   & {10}  & {7.967}  & {65}  &  {30.4}  &  {2.522}  &  {-}\\
 \hline
 {\#H8}   &  {33}   &  {10}  & {8.505}  & {40}  &  {36.3}  &  {2.524}  &  {$2.30\pm0.03$}\\
  \hline
 {\#3}   &  {34}   &  {50}  &  {8.585}  & {16}  &  {21.1}  &  {3.186}  &  {-}\\
 \hline
 {\#H6}   &  {35}   &  {10}  &  {8.519}  & {29}  &  {20.9}  &  {3.432}  &  {-}\\
 \hline
 {\#4}   &  {48}   &  {10}  & {8.970}  & {16}  &  {67.2}  &  {4.324}  &  {-}\\
 \hline
 {\#H3}   &  {50}  &  {10}  &  {8.630}  & {20}  &  {21.1}  &  {3.074}  &  {-}\\
 \hline
 {\#5}   &  {62}  &  {50}  & {8.927}  & {10} &  {57.9}  &  {4.103}  &  {-}\\
 \hline
 {\#6}   &  {80}  &  {50}  &  {9.133}  & {11}  &  {20.7}  &  {4.917}  & {-}\\ 
  \hline
 {\#H12}   &  {92}  &  {10}  &  {-}  & {-}  &  {-}  &  {-}  & {$2.12\pm0.016$}\\ 
 \hline
 \end{tabular}
\caption{The experimentally measured properties of different Nb nanofilms. From left to right: Sample name; Film thickness; width of the sample used for the electrical characterisation; $T_C$; Superconducting transition width, $\Delta T_C$; Resistivity at 300 K; Resistivity ratio (RR) at 300 K and 10 K; coherence length in the film plane at 0 K. For resistivity measurements, 1 $\mu$A of current intensity has been sourced to all films but $\#7$ (50 $\mu$A); $\#5$ (10 $\mu$A and 50 $\mu$A) and 
$\#6$ (50 $\mu$A).}
 \label{Table}
\end{table}

\begin{figure}[ht]
\renewcommand{\thefigure}{S\arabic{figure}}
\centering
\includegraphics[scale=0.55]{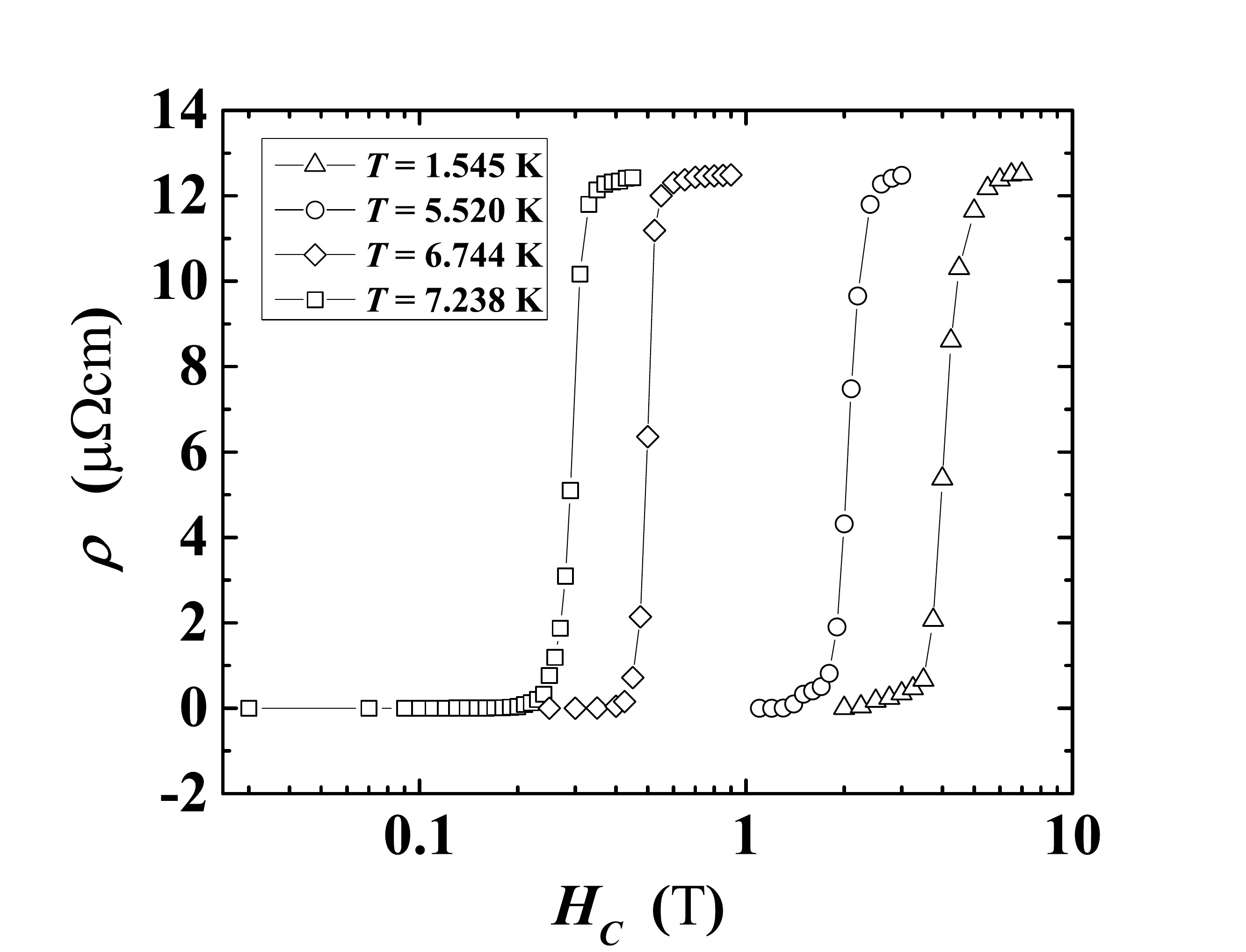}
\caption{Applied magnetic field dependence of the resistivity, at several temperatures, for a Nb film of 17 nm of thickness (film $\#H1$). Lines are a B-spline interpolation of the measured values.}  
\label{rho_H}
\end{figure}

\section*{}

\end{document}